\documentstyle[aps,epsfig,twocolumn]{revtex}

\title{Spin interactions and switching
in vertically tunnel-coupled quantum dots}

\author{
Guido Burkard$^{*}$,
Georg Seelig,
and Daniel Loss$^{\dagger}$}

\address{
Department of Physics and Astronomy,
University of Basel,\\ Klingelbergstrasse 82,
CH-4056 Basel, Switzerland}

\newcommand{\br}{{\bf r}}

\begin{document}

\twocolumn[\hsize\textwidth\columnwidth\hsize\csname @twocolumnfalse\endcsname

\maketitle

\begin{abstract}
We determine the spin exchange coupling $J$ between two electrons
located in two vertically tunnel-coupled quantum dots, and its variation
when magnetic ($B$) and electric ($E$) fields (both in-plane and
perpendicular) are applied.
We predict a strong decrease of $J$ as the in-plane
$B$ field is increased, mainly due to orbital compression.
Combined with the Zeeman splitting, this leads to
a singlet-triplet crossing, which can be observed as a pronounced jump in
the magnetization at in-plane fields of a few Tesla, and
perpendicular fields of the order of 10 Tesla for typical
self-assembled dots. We use harmonic potentials
to model the confining of electrons, and
calculate the exchange $J$ using the Heitler-London and
Hund-Mulliken technique, including the long-range Coulomb
interaction. With our results we provide experimental criteria
for the distinction of singlet and triplet states and therefore
for microscopic spin measurements.
In the case where dots of different sizes are coupled, we
present a simple method to switch on and off the spin coupling
with exponential sensitivity using an in-plane electric field.
Switching the spin coupling is essential for
quantum computation using electronic spins as qubits.
\end{abstract}


\vskip2pc]
\narrowtext

\section{Introduction}
\begin{figure}
\centerline{\psfig{file=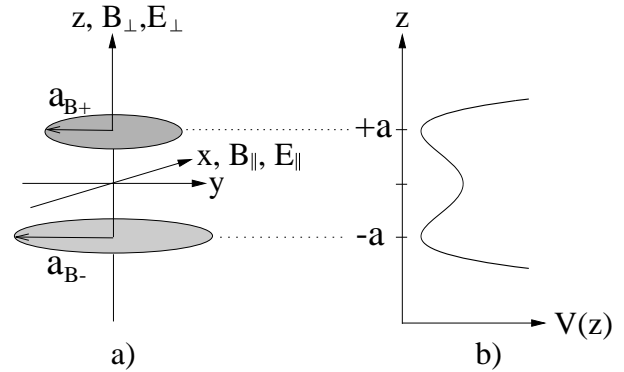,width=8cm}}
\vspace{3mm}
\caption{\label{potential}
(a) Sketch of the vertically coupled double quantum-dot
system. The two dots may have different lateral diameters,
$a_{B+}$ and $a_{B-}$. We consider magnetic and electric fields applied
either in-plane ($B_\parallel$, $E_\parallel$) or perpendicularly
($B_\perp$, $E_\perp$).
(b) The model potential for the vertical confinement
is a double well, which is obtained by combining two
harmonic wells at $z=\pm a$.
}
\end{figure}

Several methods to manipulate electronic spin in nanoscale semiconductor
devices are being developed or are already available\cite{prinz}.
Perhaps even more challenging is the proposal to use the electron spin
in quantum dots as the basic information carrier (the qubit) in a
quantum computer\cite{loss}.
The recently measured long decoherence times in semiconductor
heterostructures\cite{kikkawa} and quantum dots\cite{gupta} are
encouraging for the further research of solid-state quantum computation.
Quantum logic gates between these qubits are effected by allowing
the electrons to tunnel between two coupled quantum dots,
thereby creating an effective spin-spin interaction. There is a
large interest in quantum computation due to its potential of
solving some classically intractable problems, such as factoring\cite{shor},
and speeding up the solution of other important problems,
e.g. database search\cite{grover}. For the application of coupled quantum
dots as a quantum gate, it is important that the coupling
between the spins can be switched on and off via externally controlled
parameters such as gate voltages and magnetic fields. In a recent
publication \cite{burkard}, we have calculated the spin interaction for
two laterally coupled and identical semiconductor quantum dots defined in a
two-dimensional electron system (2DES) as a
function of these external parameters and have found that the interaction
$J$ can be switched on and off with exponential sensitivity by
changing the voltage of a gate located in between the coupled dots or
by applying a homogeneous magnetic field perpendicular to the 2DES.
In this paper, we consider
a different set-up consisting of two {\em vertically} coupled 
quantum dots with magnetic as well as electric fields applied in the plane {\em and}
perpendicular to the plane of the substrate (see Fig.~\ref{potential}).
We also extend our previous analysis to coupled quantum dots of
{\em different} sizes, which has important consequences for switching the
spin interaction: When a small dot is coupled to a large one, the
exchange coupling can be switched on and off with exponential
sensitivity using an in-plane electric field $E_\parallel$.

Semiconductor quantum dots are small engineered structures which can host
single or few electrons in a three-dimensionally confined region.
Various techniques for manufacturing quantum dots and methods for probing
their physical properties (such as electronic spectra and conductance)
are known \cite{kastner,ashoori,kouwenhoven,jacak}.
In lithographically defined quantum dots, the confinement is
obtained by electrical gating applied to a 2DES
in a semiconductor heterostructure, e.g. in AlGaAs/GaAs.
In vertical dots, a columnar mesa structure is produced by etching
a semiconductor heterostructure\cite{tarucha}.
While laterally coupled quantum dots have been defined in 2DES by tunable
electric gates\cite{waugh,livermore,blick,oosterkamp98a},
vertically coupled dots have been manufactured either by etching a mesa
structure out of a triple-barrier heterostructure and subsequently placing
an electrical side-gate around it\cite{austing} or by using stacked
double-layer self-assembled dots (SADs) \cite{fafard,luyken}. In the
mesa structure, the number of electrons per dot can be varied one by one
starting from zero, whereas in SADs the average number of electrons
per dot in a sample with many dots can be controlled, even one electron
per dot is experimentally feasible \cite{fricke}.  

Self-assembled quantum dots are manufactured in the so-called
Stranski-Krastanov growth mode where
a lattice-mismatched semiconducting material is epitaxially grown on a
substrate, e.g. InAs on GaAs\cite{sads}.
Minimization of the lattice mismatch strain occurs through the formation
of small three-dimensional islands.
Repeating the fabrication procedure described above, a second 
layer of quantum dots can be formed on top of the first one. Since 
the strain field of a dot in the first layer acts as a nucleus for the 
growth of a dot in the second layer, the quantum dots in the two layers are 
strongly spatially correlated\cite{verticalsads}. 
Electrostatic coupling in vertical SADs has been investigated \cite{luyken},
and it can be expected that the production of tunnel-coupled SADs will 
be possible in the near future.

In this paper, we concentrate on the magnetic properties (including
in-plane fields, $B_\parallel$) of pairs of quantum dots in which two
electrons are vertically coupled via quantum tunneling and are subject
to the full Coulomb interaction. See Fig.~\ref{potential} for a sketch
of the system under study.
Coupled quantum dots in the absence of quantum tunneling (purely
electrostatic interactions) were studied in
Refs.~\onlinecite{palacios,benjamin,partoens}.
Electronic spectra and charge densities for two electrons in a system
of vertically tunnel-coupled quantum dots at zero magnetic field were
calculated in Ref.~\onlinecite{bryant}.
Singlet-triplet crossings in the ground state of single\cite{wagner} and
coupled dots with two\cite{oh} to four\cite{imamura,tokura} electrons in
vertically coupled dots in the presence of a magnetic field perpendicular
to the growth direction ($B_\perp$ in Fig.~\ref{potential}) have been
predicted.

In contrast to previous theoretical work on coupled dots
\cite{palacios,benjamin,partoens,bryant,wagner,oh,imamura,tokura}
the investigation presented here both takes into account quantum tunneling
and includes \textit{in-plane} magnetic fields
($B_\parallel$ in Fig.~\ref{potential}),
leading to a much stronger suppression of the
exchange energy than for $B_\perp$ (for very weakly confined dots, in-plane
$B$ fields can cause a singlet-triplet crossing, even in the absence of the
Zeeman coupling). This result is in analogy to
our earlier finding of a spin singlet-triplet crossing in laterally
coupled identical dots as the perpendicular field is increased\cite{burkard}.
In addition to this, we investigate the influence of an electric field
$E_\perp$
applied in growth direction on the low-energy electronic levels in the
vertically coupled quantum dots. From the electronic spectrum, we derive
the equilibrium magnetization as a function of both the magnetic and the
electric fields (magnetization measurements for many-electron double
quantum dots are reported in Ref.~\onlinecite{oosterkamp98b}).
As another important extension of earlier work, we consider a small dot
which is tunnel-coupled to a large dot. We find that this system
represents an ideal candidate for a quantum gate, since the exchange
interaction $J$ can be switched simply by applying an in-plane
electric field $E_\parallel$ (see Sec.~\ref{e_switch}).

Our main interest is in the dynamics of the spins of the two electrons
which are confined in the double dot. The spin dynamics can be
described by an isotropic Heisenberg interaction, 
\begin{equation}\label{Heisenberg}
H_{\rm s}=J\,\,{\bf S}_1\cdot{\bf S}_2,
\end{equation}
where the exchange energy $J$ is the difference of the energies of 
the two-particle ground-state, a spin-singlet at zero magnetic field, and 
the lowest spin-triplet state. We shall calculate the 
exchange energy $J({\bf B},{\bf E},a)$ of two vertically
coupled quantum dots containing one electron each as a function of
electric and magnetic fields (${\bf E}$ and ${\bf B}$) and the inter-dot
distance $2a$.
We show that an in-plane magnetic field has a much stronger influence on
the spin coupling than a perpendicular magnetic field.
Moreover, we will discuss the influence of the dot size on $J$, and 
investigate systems containing two dots of  different sizes.
We will see that it is possible to suppress the spin-spin coupling
exponentially by means of an in-plane magnetic field $B_\parallel$ for
large dots (weak confinement) or, alternatively, with an in-plane
electric field $E_\parallel$ if one of the dots is larger than the other.
Furthermore we will point out differences and similarities in the 
field-dependence of the tunnel-splitting $t$ found in a quantum mechanically 
coupled double-dot system containing only a single electron and the exchange 
energy $J$, a quantity due to two-particle correlations.
Performing these 
calculations we make use of methods known from molecular physics 
(Heitler-London and Hund-Mulliken technique) thus exploiting the analogy 
between quantum dots and atoms. Note again that besides being interesting 
in its own right, a quantum-dot ``hydrogen molecule'', if experimentally 
controllable, could be used as a fundamental part of a solid-state 
quantum-computing device \cite{loss,burkard}, using the electronic
spin as the qubit.

In our discussion of the 
vertically coupled double-dot system we proceed as follows.
In Section \ref{model} we introduce a model for the description
of a vertical double-dot structure. Subsequently (Sec.~\ref{perpendicular}),
we discuss vertically coupled quantum dots in perpendicular magnetic
and electric fields.  Sec.~\ref{parallel} is devoted to the discussion of a
double-dot structure in the presence of
an in-plane magnetic field. In Sec.~\ref{e_switch}
we present a simple switching mechanism for the spin coupling involving
an in-plane electric field. Finally, we discuss the implications of
our result for two-spin and single-spin measurements in
Sec.~\ref{spin_meas}.

\section{Model}\label{model}

The Hamiltonian which we use for the description of two vertically coupled 
quantum dots is
\begin{eqnarray}\label{H}
H &=& \sum_{i=1,2} h({\bf r}_i, {\bf p}_i)+C, \nonumber\\
h({\bf r}, {\bf p}) &=& \frac{1}{2m}\left({\bf p}-\frac{e}{c}{\bf A}(\br) 
\right)^2+ezE+V_l(\br)+V_v(\br),\label{hamiltonian}\\ 
C&=&\frac{e^2}{\kappa\left| \br_1-\br_2\right|},\nonumber
\end{eqnarray} 
where $C$ is the Coulomb interaction and $h$ the single-particle 
Hamiltonian.  The dielectric constant $\kappa$ and the effective mass 
$m$ are material parameters.  The potential $V_l$ in $h$ describes the lateral 
confinement, whereas $V_v$ models the vertical double-well structure.  For 
the lateral confinement we choose the parabolic potential 
\begin{equation}\label{lateral}
V_l (x,y)=\frac{m}{2}\omega^2_{z}
\left\{
\begin{array}{l l}
\alpha^2_{0+} (x^2 + y^2), & z>0,\\
\alpha^2_{0-} (x^2 + y^2), & z<0,\\
\end{array}
\right.
\end{equation}
where we have introduced the anisotropy parameters
$\alpha_{0\pm}$ determining the strength of the 
vertical relative to the lateral confinement. Note that for dots of
different size ($\alpha_{0+}\neq\alpha_{0-}$) the model
potential Eq.~(\ref{lateral}) is not continuous at  $z=0$. 
The lateral effective Bohr radii
$a_{B\pm}=\sqrt{\hbar/(m\omega_z\alpha_{0\pm}})$ 
are a measure for the lateral extension of the electron wave function in 
the dots.  In experiments with electrically gated quantum dots 
in a two-dimensional electron system (2DES), it has been shown that the 
electronic spectrum is well described by a simple harmonic oscillator 
\cite{kouwenhoven,jacak}.
In the presence of a magnetic field $B_\perp$ perpendicular to the 2DES,
the one-particle problem has the Fock-Darwin states\cite{fock} as an
exact solution. Furthermore, it has been shown 
experimentally \cite{fricke} and theoretically \cite{hawrylak} that a 
two-dimensional harmonic confinement potential is a reasonable 
approximation to the real confinement potential in a lens-shaped SAD.
In describing the  confinement $V_v$ along the inter-dot axis,
we have used a (locally harmonic) double well potential of the form
(see Fig.~\ref{potential}b)
\begin{equation}\label{VOsc}
V_v = \frac{m\omega_z^2}{8a^2}\left(z^2-a^2 \right)^2,
\end{equation}
which, in the limit of large inter-dot distance $a\gg a_{\rm B}$,
separates (for $z\approx \pm a$) into two harmonic wells (one for each dot) 
of frequency $\omega_z$. Here $a$ is half the distance between the centers
of the dots and $a_B=\sqrt{\hbar/(m\omega_z})$ is the 
vertical effective Bohr radius.
For most vertically coupled dots, the vertical 
confinement is determined by the conduction band offset between different 
semiconductor layers; therefore in principle a square-well potential would
be a more accurate description of the real potential than the harmonic
double well (note however, that the required conduction-band offsets
are not always known exactly).
There is no qualitative difference between the results presented below
obtained with harmonic potentials and the corresponding
results which we obtained using square-well potentials\cite{seelig}.

It was shown in Refs.~\cite{burkard,burkard2} that the spin-orbit contribution (due to the confinement)
$H_{\rm so}=\left(\omega^2_z/2m_ec^2\right){\bf S}\cdot{\bf L}$ with $m_e$
being the bare electron mass can be neglected in the relevant cases, e.g.
$H_{\rm so}/\hbar\omega_z\sim 10^{-7}$ for $\hbar\omega_z=30\,{\rm meV}$
in GaAs.

The Zeeman-splitting $H_Z=g\mu_B\sum_{i=1,2}{\bf B}\cdot{\bf S}_i$ is not 
included in the two-particle Hamiltonian Eq.~(\ref{H}), since in the
absence of spin-orbit coupling one can treat the orbital problem separately
and include the Zeeman interaction later (which we will do when we study the
low-energy spectra and the magnetization).
Here, we have denoted the effective g-factor with $g$ and the Bohr magneton
with $\mu_B$.

\section{Perpendicular Magnetic Field $B_\perp$}\label{perpendicular}

We first study the vertically coupled double dot in a perpendicular
magnetic field $B=B_\perp$ (cf. Fig.~\ref{potential})
which corresponds to the vector potential
${\bf A}({\bf r})= {\rm B}(-y,x,0)/2$
in the symmetric gauge (for the time being, we set $E=0$).

The confining potentials for the two electrons are given 
in Eqs.~(\ref{lateral}) and (\ref{VOsc}).  As a starting point for our 
calculations we consider the problem of an electron in a single quantum dot.  
The one-particle Hamiltonian by which we describe a single electron in the 
upper (lower) dot of the double-dot system is
\begin{eqnarray}
h^0_{\pm a}({\bf r})
&=&\frac{1}{2m}\left({\bf p}-\frac{e}{c}{\bf A}({\bf r})\right)^2\nonumber\\
&&+\frac{m\omega^2_z}{2}\left(\alpha_{0\pm}^2(x^2+y^2)+(z\mp a)^2\right),
\label{h0}
\end{eqnarray}
and has the ground-state Fock-Darwin\cite{fock} solution
\begin{equation}\label{oneparticle}
\varphi_{\pm a}({\bf r}) = \left(\frac{m \omega_z}{\pi\hbar}\right)^{3/4}
\sqrt{\alpha_\pm}
e^{-\frac{m\omega_z}{2\hbar}\left (\alpha_\pm \left(x^2+ y^2\right)
+(z \mp a)^2\right)}, 
\end{equation}
corresponding to the ground-state energy
$\epsilon_{\pm}=\hbar\omega_z(1+2\alpha_{\pm})/2$.
In Eq.~(\ref{oneparticle}) we have introduced $\alpha_{\pm}({\rm B}) 
=\sqrt{\alpha^2_{0\pm}+\omega_L(B)^2/\omega^2_z}
=\sqrt{\alpha^2_{0\pm}+B^2/B^2_0}$,
with $\omega_L(B) =eB/2mc$ the Larmor frequency and $B_0=2mc\omega_z/e$ 
the magnetic field for which $\omega_z=\omega_L$.  The parameters 
$\alpha_{\pm}(B)$ describe the compression of the one-particle wave 
function perpendicular to the magnetic field.  For finding the exchange 
energy $J$ we make the Heitler-London ansatz, using the symmetric and 
antisymmetric two-particle wave-functions
$|\Psi_{\pm}\rangle = (|12\rangle \pm |21\rangle)/\sqrt{2(1\pm S^2)}$, where we
use the one-particle orbitals $\varphi_{-a}({\bf r})=\langle {\bf 
r}|1\rangle$ and $\varphi_{+a}({\bf r})=\langle {\bf r}|2\rangle$.
Here $|ij\rangle = |i\rangle |j\rangle$ are two-particle product states and
$S=\int d^3r\varphi_{+a}^{*}(\br)\varphi_{-a}(\br)=\langle 2|1\rangle$ denotes
the overlap of the right and left orbitals.  A non-vanishing overlap $S$ implies 
that the electrons can tunnel between the dots.  Using the two-particle 
orbitals $|\Psi_\pm\rangle$ we can calculate the singlet and triplet energy 
$\epsilon_{\rm s/t}=\langle\Psi_{\pm}|H|\Psi_{\pm}\rangle $ and therefore
the exchange energy
$J=\epsilon_t-\epsilon_s$. We rewrite the Hamiltonian, adding and subtracting
the potential of the single upper (lower) dot for electron $1(2)$ in $H$. The 
Hamiltonian then takes the form $H=h^0_{-a}({\bf r}_1)+h^0_{+a}({\bf 
r}_2)+W+C$ which is convenient, because it 
contains the single particle Hamiltonians $h^0_{+a}$ and $h^0_{-a}$ of 
which we know the exact solutions.  The potential term 
is $W({\bf r_1},{\bf r_2})=W_l(x_1,y_1,x_2,y_2)+W_v(z_1,z_2)$, where 
\begin{eqnarray}
\lefteqn{W_l(x_1,y_1,x_2,y_2)=\sum_{i=1,2} V_l(x_i,y_i)}\nonumber\\
&&\hspace{2cm}-\frac{m\omega^2_z}{2}\left[\alpha^2_{0-}(x^2_1+y^2_1)
+\alpha^2_{0+}(x^2_2+y^2_2)\right],
\end{eqnarray}
\begin{equation}
W_v(z_1,z_2)=\sum_{i=1,2} V_{v}(z_i)
-\frac{m\omega^2_z}{2}\left[(z_1+a)^2+(z_2-a)^2\right].
\end{equation}
The formal expression for $J$ is now
\begin{equation}\label{Jformal}
J = \frac{2S^2}{1-S^4}\left(\langle 12|C+W|12\rangle-\frac{{\rm Re} \langle 
12|C+W|21\rangle}{S^2}\right).
\end{equation}
Evaluation of the matrix elements $\langle 12|C+W|12\rangle$ and 
$\langle 12|C+W|21\rangle$ provides us with the result,
\begin{eqnarray}\label{JHLBz}
J &=& 
\frac{2S^2}{1-S^4}\hbar\omega_z \bigg[ c \sqrt{\mu}\,e^{2\mu 
d^{2}}\left(1-{\rm erf}\left(d \sqrt{2\mu}\right)\right) \nonumber \\
&-& \frac{c}{\pi} \frac{\alpha_+ + \alpha_-}{\sqrt{1-(\alpha_+ + \alpha_- 
-1)^{2}}}\, {\rm arccos}(\alpha_+ + \alpha_- -1) \\
&+&  \frac{1}{4} \left(\alpha^2_{0+}-\alpha^2_{0-}\right)\frac{\alpha_+ - 
\alpha_-}{\alpha_+ \alpha_-}\, \left(1-{\rm 
erf(d)}\right)+\frac{3}{4} \left(1+d^2\right) \bigg],\nonumber
\end{eqnarray}
where ${\rm erf}(x)$ denotes the error function.
We have introduced the dimensionless parameters $d=a/a_B$ for the inter-dot 
distance, and $c=\sqrt{\pi/2}(e^2/\kappa a_{\rm B})/\hbar\omega_z$ for the
Coulomb interaction. Note that
$\alpha_{\pm}$, $\mu=2\alpha_+ \alpha_-/\left(\alpha_+ +\alpha_-\right)$,
and the overlap
\begin{equation}\label{overlaposc} 
S = 2\frac{\sqrt{\alpha_{+}\alpha_{-}}}{\alpha_{+} 
+\alpha_{-}}\,{\rm exp}(-d^2),
\end{equation}
depend on the magnetic field $B$.
The first term in the 
square brackets in Eq.~(\ref{JHLBz}) is an approximate evaluation of the 
direct Coulomb integral $\langle 12|C|12\rangle$ for $d\gtrsim 0.7$ and
for magnetic fields $B\lesssim B_0$\onlinecite{footnote1}.
The second term in Eq.~(\ref{JHLBz}) is the (exact) exchange Coulomb integral
$\langle 12|C|21\rangle/S^2$, while the last two terms stem from the potential 
integrals, which were also evaluated exactly.  If the two dots have the 
same size, the expression for the exchange energy Eq.~(\ref{JHLBz}) can be 
simplified considerably. We will first study the case of two dots of 
equal size, and later come back to the case of dots which differ in size.

Setting $\alpha_{0+}=\alpha_{0-}\equiv\alpha_0$ in 
Eq.~(\ref{JHLBz}) and using Eq.~(\ref{overlaposc}), we obtain 
\begin{eqnarray}\label{Jequal}
J &=& \frac{\hbar\omega_z}{{\rm sinh}(2d^2)}
\Bigg[ c \sqrt{\alpha} e^{2\alpha d^{2}}
\left(1-{\rm erf}\left(d \sqrt{2 \alpha}\right)\right)\\ 
&-& \frac{c}{\pi} \frac{2\alpha}{\sqrt{1-(2\alpha -1)^{2}}}
\arccos (2\alpha -1)+\frac{3}{4} \left(1+d^2\right) \Bigg],\nonumber 
\end{eqnarray} 
where $\alpha=\sqrt{\alpha^2_0+B^2/B^2_0}$.  
As before, the first term in Eq.~(\ref{Jequal}) is the direct Coulomb term,
while the second term (appearing with a negative sign) is the exchange
Coulomb term. Finally, the potential term in this case equals $W=(3/4)(1+d^2)$
and is due to the vertical confinement only.
For two dots of equal size neither the prefactor $2S^2/(1-S^4)$ nor 
the potential term depends on the magnetic field.  Since the direct Coulomb
term depends on $B_\perp$ only weakly, the field dependence 
of the exchange energy is mostly determined by the exchange Coulomb term.  

Note that for obtaining the large-field asymptotics ($B\gtrsim B_0$)
it would be necessary to include
hybridized one-particle wave functions\cite{burkard} since in the magnetic
field the level spacings between the one-particle ground states are shrinking
and eventually become smaller than $J$, thus undermining the self-consistency
of the one-orbital Heitler-London approximation.    
Increasing the inter-dot distance $d$ (for fixed confinement $\hbar\omega$),
an exponential 
decrease of the exchange energy $J$ is predicted by Eqs.~(\ref{JHLBz}) and 
(\ref{Jequal}). As mentioned, Eq.~(\ref{JHLBz}) is an approximation and 
should not be used for small inter-dot distances $d\lesssim 0.7$.  There 
are also some limitations to the choice of the anisotropy parameters 
$\alpha_{0 \pm}$.  If we consider a system with much stronger vertical than 
lateral confinement (e.g.  $\alpha_{0\pm}=1/10$), the exchange energy will 
become larger than the smallest excitation energy 
$\Delta\epsilon=\alpha_{0\pm}\hbar\omega_z$ in the single dot spectrum.  In 
that case we have to improve our Heitler-London approach by including 
hybridized single-dot orbitals\cite{burkard}.  If, on the other hand, the 
two dots are different in size, double occupation of the larger dot is 
energetically favorable and a Hund-Mulliken approach should be employed.
In the Hund-Mulliken approximation, the Hilbert space for the spin singlet is
enlarged by including two-particle states describing double occupation of
a quantum dot.  Since only the singlet sector is enlarged it can be expected
that we obtain a lower singlet energy $\epsilon_{\rm s}$ than from 
the Heitler-London calculation (but the same triplet energy 
$\epsilon_{\rm t}$) and therefore $J=\epsilon_{\rm t}-\epsilon_{\rm s}$ will
be larger than the Heitler-London result, Eq. (\ref{JHLBz}).

\begin{figure}
\begin{tabular}{r r r}
\hspace{-4mm}\psfig{file=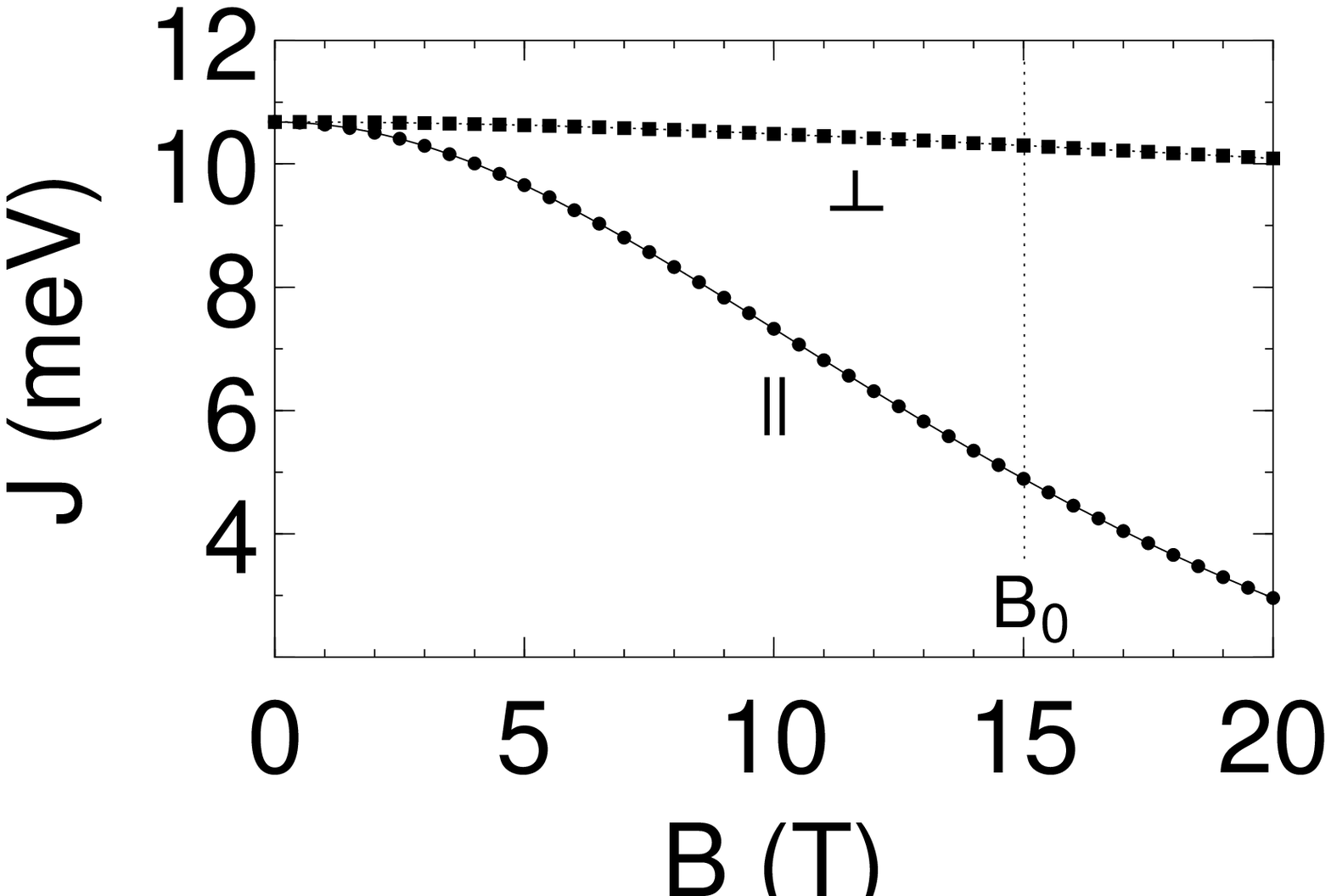,scale=0.24} &
&
\psfig{file=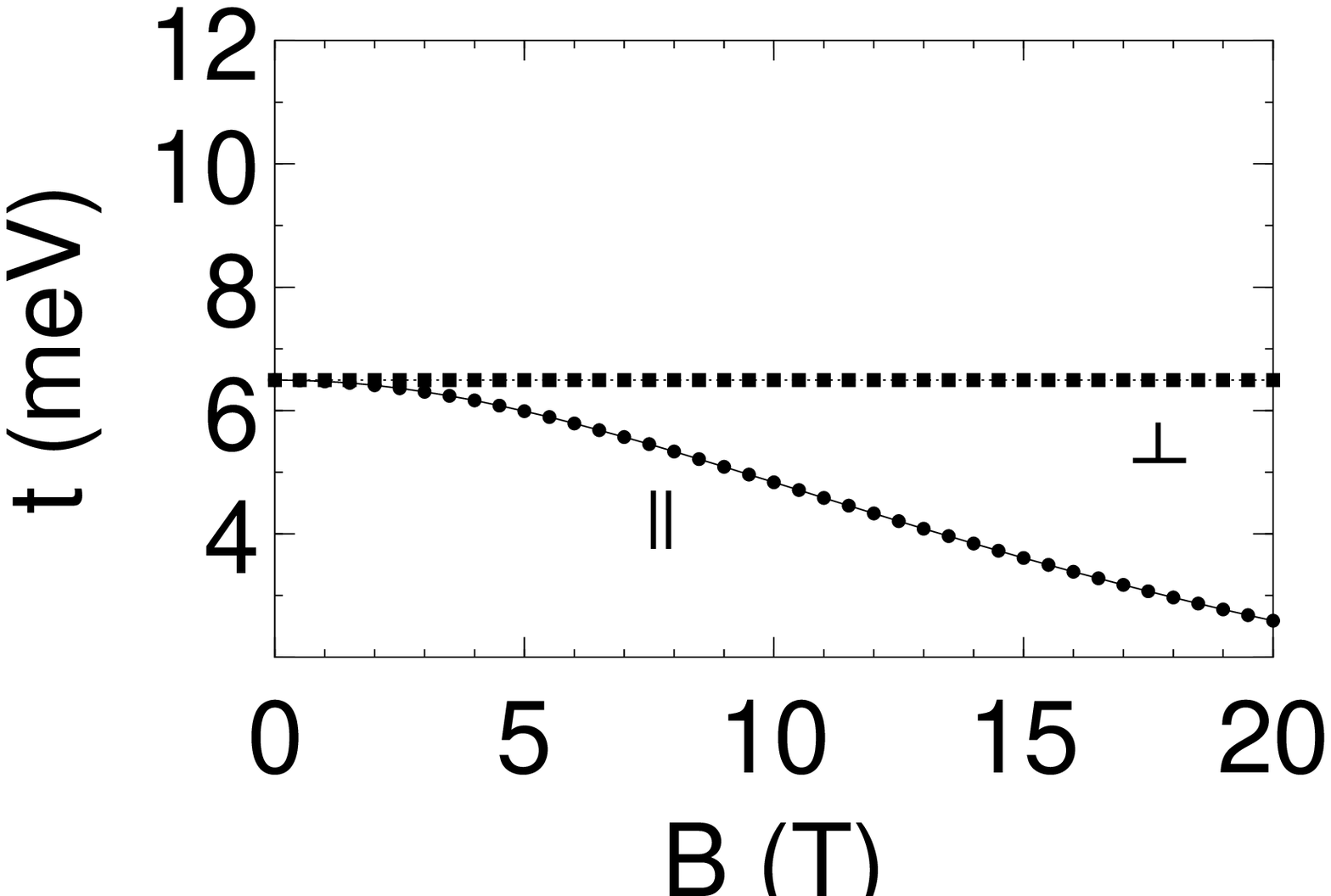,scale=0.24}
\end{tabular}
\caption{\label{osc_fig_1b}
Exchange energy $J$ (left graph) and single-electron tunneling amplitude
$t$ (right graph) as a function of the applied magnetic field for
two vertically coupled small (height $6\,{\rm nm}$, width $12\,{\rm nm}$)
InAs ($m=0.08m_e$, $\kappa=14.6$)
quantum dots (e.g. self-assembled dots) in a center-to-center
distance of $9\,{\rm nm}$ ($d=1.5$). The box-shaped symbols correspond
to the magnetic field $B_\perp$ applied in $z$ direction, the circle
symbols to the field $B_\parallel$ in $x$ direction.
The plotted results were obtained using the Hund-Mulliken method and
are reliable up to a field $B_0\approx 15\,{\rm T}$ where
higher levels start to become important.
}
\end{figure}

We now apply the Hund-Mulliken approach to calculate the exchange energy 
of the double-dot system.
We therefore introduce the orthonormalized one-particle wave functions
$\Phi _{\pm a}=(\varphi_{\pm a}-g\varphi_{\mp a})/\sqrt{1-2Sg+g^2}$,
where $g=(1-\sqrt{1-S^2})/S$. Using $\Phi_{\pm a}$, we generate four
basis functions with respect to which we diagonalize the two-particle
Hamiltonian $H$: The states with double occupation,
$\Psi^{\rm d}_{\pm a}(\br_1,\br_2)=\Phi_{\pm a}(\br_1)\Phi_{\pm a}(\br_2)$
and the states with single occupation $\Psi^{\rm s}_{\pm}(\br_1,\br_2)
=[\Phi_{+a}(\br_1)\Phi_{-a}(\br_2)\pm \Phi_{-a}(\br_1)\Phi_{+a}(\br_2)]
/\sqrt{2}$.
Calculating the matrix elements of the Hamiltonian $H$ in this orthonormal
basis we find 
\begin{equation}\label{matrix}
H =\left (\begin{array}{cccc}
2\epsilon + V_+&-\sqrt{2}t_{{\rm H}+}&-\sqrt{2}t_{{\rm H}-}&0\\ -\sqrt{2}t_{{\rm 
H}+}&2\epsilon_+ +U_+&X&0\\ -\sqrt{2}t_{{\rm H}-}&X&2\epsilon_- +U_-&0\\ 
0&0&0&2\epsilon + V_-\\ \end{array}\right),
\end{equation}
where
\begin{eqnarray}
\epsilon_{\pm} &=& 
\langle\Phi_{\pm a}|h(z\mp a)|\Phi_{\pm a}\rangle ,\quad
\epsilon = \frac{1}{2}\left(\epsilon_++\epsilon_-\right),
\label{epsilon}\\
t_{{\rm H}\pm} &=& t- w_{\pm}
= -\langle\Phi_{\pm a}|h|\Phi_{\mp a}\rangle
- \frac{1}{\sqrt{2}}\langle\Psi^{\rm s}_+|C|\Psi^{\rm d}_{\pm a}\rangle
,\label{t}\\ 
V_{\pm} &=& \langle\Psi^{\rm s}_{\pm}|C|\Psi^{\rm s}_{\pm}\rangle,
\quad\quad
U_{\pm} =\langle\Psi^{\rm d}_{\pm a}|C|\Psi^{\rm d}_{\pm a}\rangle,
\label{C_diag}\\
X &=& \langle \Psi^{\rm d}_{\pm a}|C|\Psi^{\rm d}_{\mp a}\rangle .
\label{C_offdiag}
\end{eqnarray}
\begin{figure}
\begin{tabular}{r r r}
\hspace{-4mm}\psfig{file=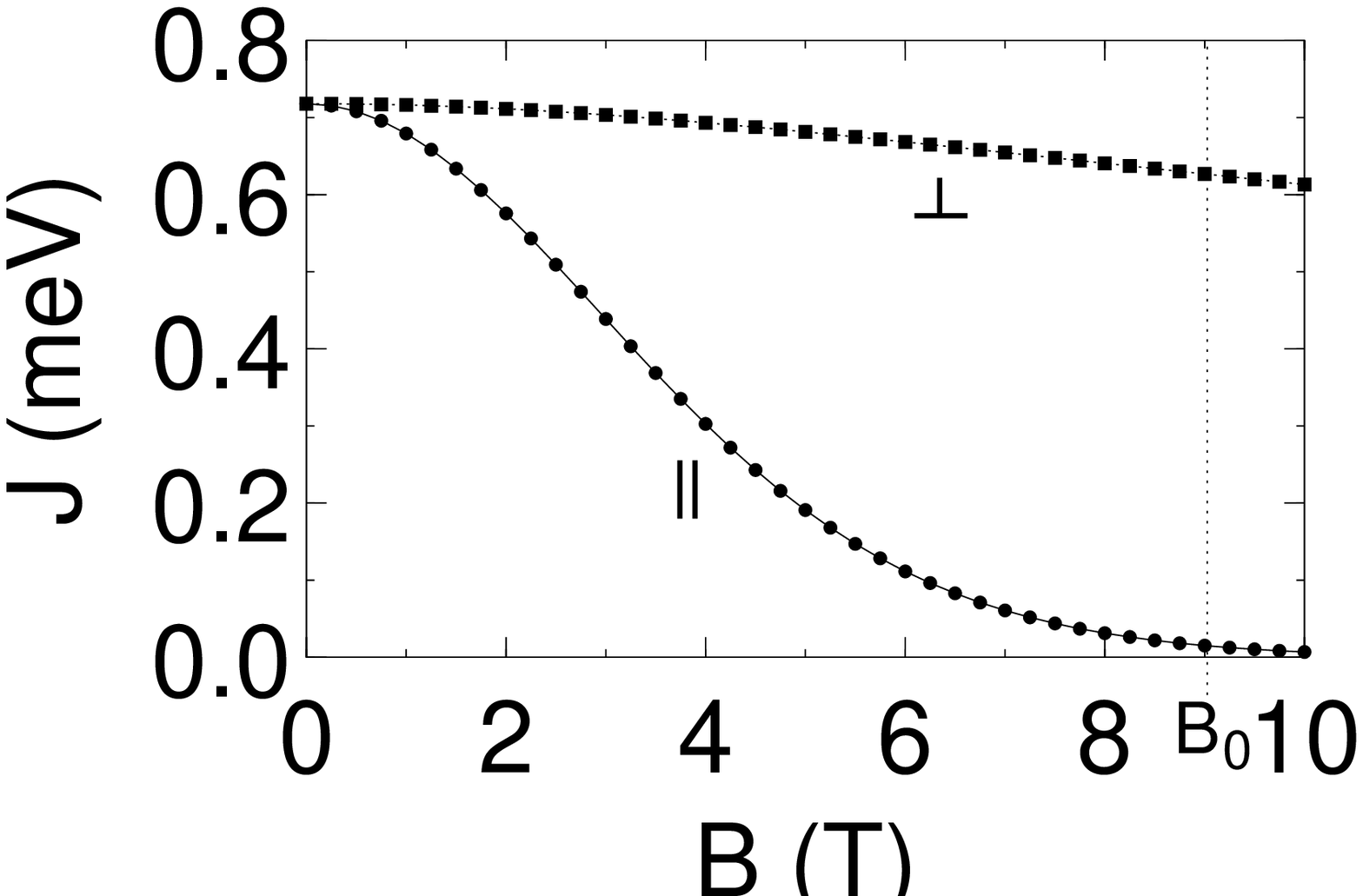,scale=0.24} &  &
\psfig{file=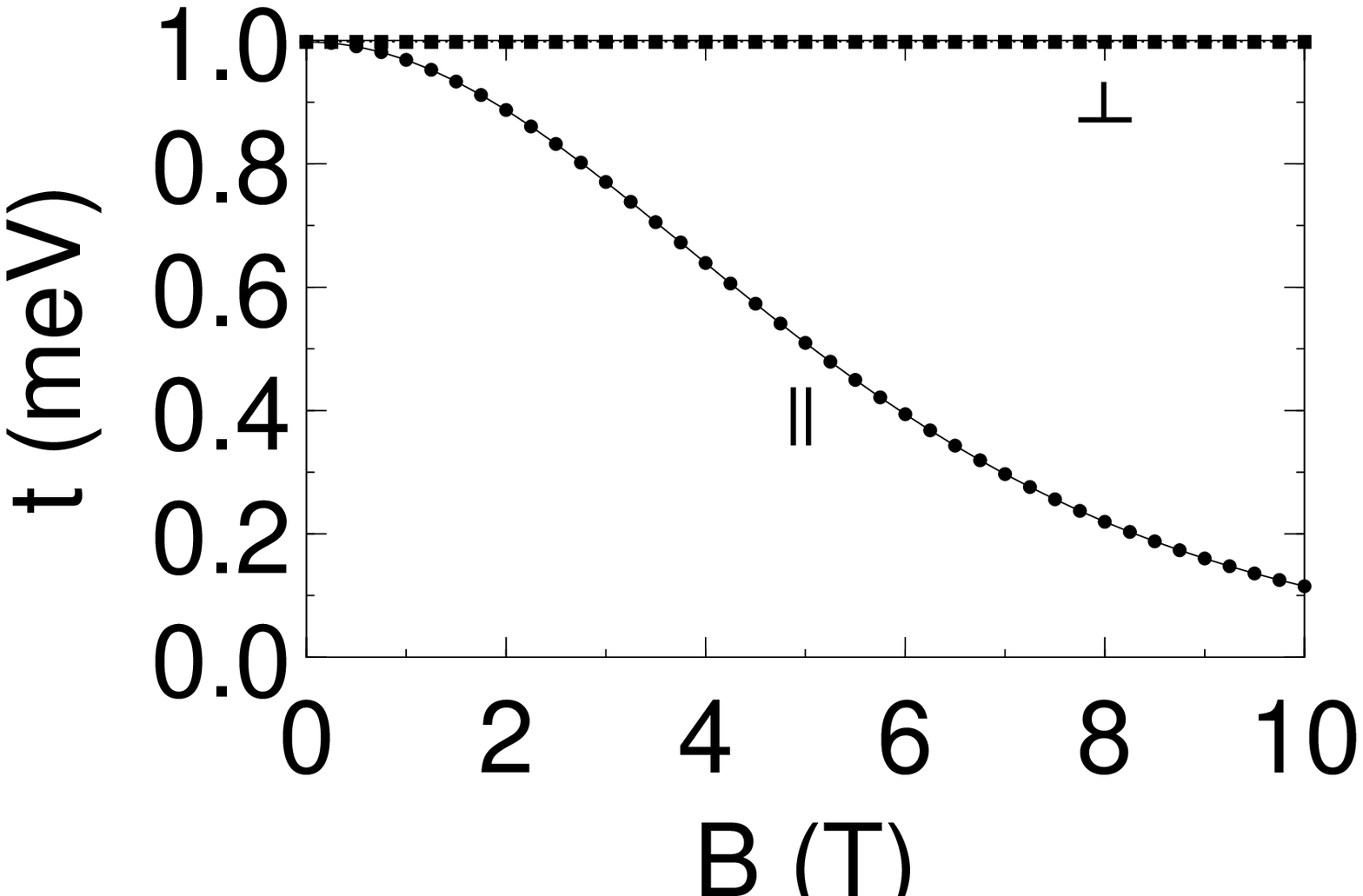,scale=0.24}
\end{tabular}
\caption{\label{osc_fig_1}
Left graph:
Exchange energy $J$ as a function of the magnetic field $B$ applied
vertically to the $xy$ plane ($B_\perp$, box symbols) and in-plane
($B_\parallel$, circle symbols), as 
calculated using the Hund-Mulliken method.
Note that due to vertical orbital compression,
the exchange coupling decreases much more
strongly for an in-plane magnetic field.
The parameters for this plot correspond to a system of two equal
GaAs dots,  each $17\,{\rm nm}$ high and $24\,{\rm nm}$ in diameter
(vertical confinement energy $\hbar\omega_z=16\,{\rm meV}$ and
anisotropy parameter $\alpha_0=1/2$). The dots are located at a
center-to-center distance of $2a=31\,{\rm nm}$ ($d=1.8$).
The single-orbital approximation breaks down at about
$B_0\approx 9\,{\rm T}$, where it is expected that levels which
are higher in the zero-field ($B=0$) spectrum determine the
exchange energy.
Right graph:
Single-particle tunneling amplitude $t$ vs. magnetic field
for the same system.
Note that in contrast to the exchange coupling (a genuine
two-particle quantity), $t$ describes the tunneling of
a \textit{single} particle. Whereas $J$ shows a weak dependence
on the vertical magnetic field $B_{\perp}$, we note that
$t(B_{\perp})$ (box symbols) is constant.
}
\end{figure}
The general form of the entries of the matrix Eq.~(\ref{matrix})
are given in Appendix \ref{appendix_matrixelements}.
The evaluation for perpendicular
magnetic fields $B_\perp$ can be found
in Appendix \ref{appendixHMosc}.
We do not display the eigenvalues of the matrix Eq.~(\ref{matrix}) here,
since the expressions are lengthy.  However, if the two 
dots have the same size ($\alpha_{0-}=\alpha_{0+}$), then the Hamiltonian
considerably simplifies since
$t_{{\rm H}-} = t_{{\rm H}+} \equiv t_{{\rm H}}$, $\epsilon_+ =\epsilon_- 
\equiv\epsilon$ and $U_+ = U_- \equiv U$.
In this case the eigenvalues are
$\epsilon_{{\rm s}\pm} = 2\epsilon + U_{\rm H}/2+V_+\pm\sqrt{U_{\rm 
H}^2/4+4t_{\rm H}^2}$ and $\epsilon_{{\rm s} 0} = 2\epsilon + U_{\rm H} - 
2X + V_+$ for the three singlets, and $\epsilon_{\rm t} = 2\epsilon + V_-$ 
for the triplet, where we have introduced the additional quantity $U_{\rm 
H} = U-V_+ +X$.  The exchange energy is the difference between the lowest
singlet and the triplet state,
\begin{equation}\label{HMresult}
J = \epsilon_{\rm t}-\epsilon_{{\rm s}-}
=V - \frac{U_{\rm H}}{2} + \frac{1}{2}\sqrt{U_{\rm H}^2 + 16t_{\rm H}^2},
\end{equation}
where we have used $V=V_- - V_+$.
The singlet energies $\epsilon_{{\rm s}+}$ and $\epsilon_{{\rm s}0}$ are 
separated from $\epsilon_{\rm t}$ and $\epsilon_{{\rm s}-}$ by a gap of 
order $U_{\rm H}$ and are therefore negligible for the study of low-energy 
properties.  If only short-range Coulomb interactions are considered (which
is usually done in the standard Hubbard approach) the exchange energy $J$
reduces to $-U/2 + \sqrt{U^2+ 16t^2}/2$, where $t$ and $U$ denote the hopping
matrix element and on-site repulsion which are not renormalized
by interaction. 
We call the quantities $t_{\rm H}$ and $U_{\rm H}$ the \textit{extended}
hopping matrix element and \textit{extended} on-site repulsion, 
respectively, since they are renormalized by long-range Coulomb 
interactions.  If the Hubbard ratio $t_H/U_H$ is $\lesssim 1$, we are in the
Hubbard limit, where $J$ approximately takes the form
(cf.\ Ref.~\onlinecite{burkard}) 
\begin{equation}
J = \frac{4t_{\rm H}^2}{U_{\rm H}}+V.  \label{Hubbard} 
\end{equation}
The first term in Eq.~(\ref{Hubbard}) has the form of the standard Hubbard
model result, whereas the second term $V$ is due to the long-range Coulomb 
interactions and accounts for the difference in Coulomb energy between the
singlet and triplet states $\Psi^{\rm s}_{\pm}$.
We have evaluated our result for a GaAs ($m=0.067m_e$, $\kappa=13.1$)
system comprising two equal dots with vertical confinement energy
$\hbar\omega_z=16\,{\rm meV}$ ($a_B=17\,{\rm nm}$) and horizontal confinement
energy $\alpha_0\hbar\omega_z=8\,{\rm meV}$ in a distance $a=31\,{\rm nm}$
($d=1.8$). The result is plotted in Fig.~\ref{osc_fig_1} (left graph,
box-shaped symbols).
The exchange energy $J(B_\perp)$ as obtained from
the Hund-Mulliken method for two coupled InAs SADs ($m=0.08m_e$\cite{fricke},
$\kappa=14.6$, $\hbar\omega_z=50\,{\rm meV}$,
$\alpha_{0+}=\alpha_{0-}=1/4$) is plotted in Fig.~\ref{osc_fig_1b}
(left graph, box symbols).
Including the Zeeman splitting, we can now plot the low-energy spectrum
as a function of the magnetic field, see Fig.~\ref{osc_fig_2} (left).
Note that the spectrum clearly differs from the single-electron spectrum
in the double-dot (Fig.~\ref{osc_fig_2}, right).

\begin{figure}
\begin{tabular}{r r r}
\hspace{-4mm}
\psfig{file=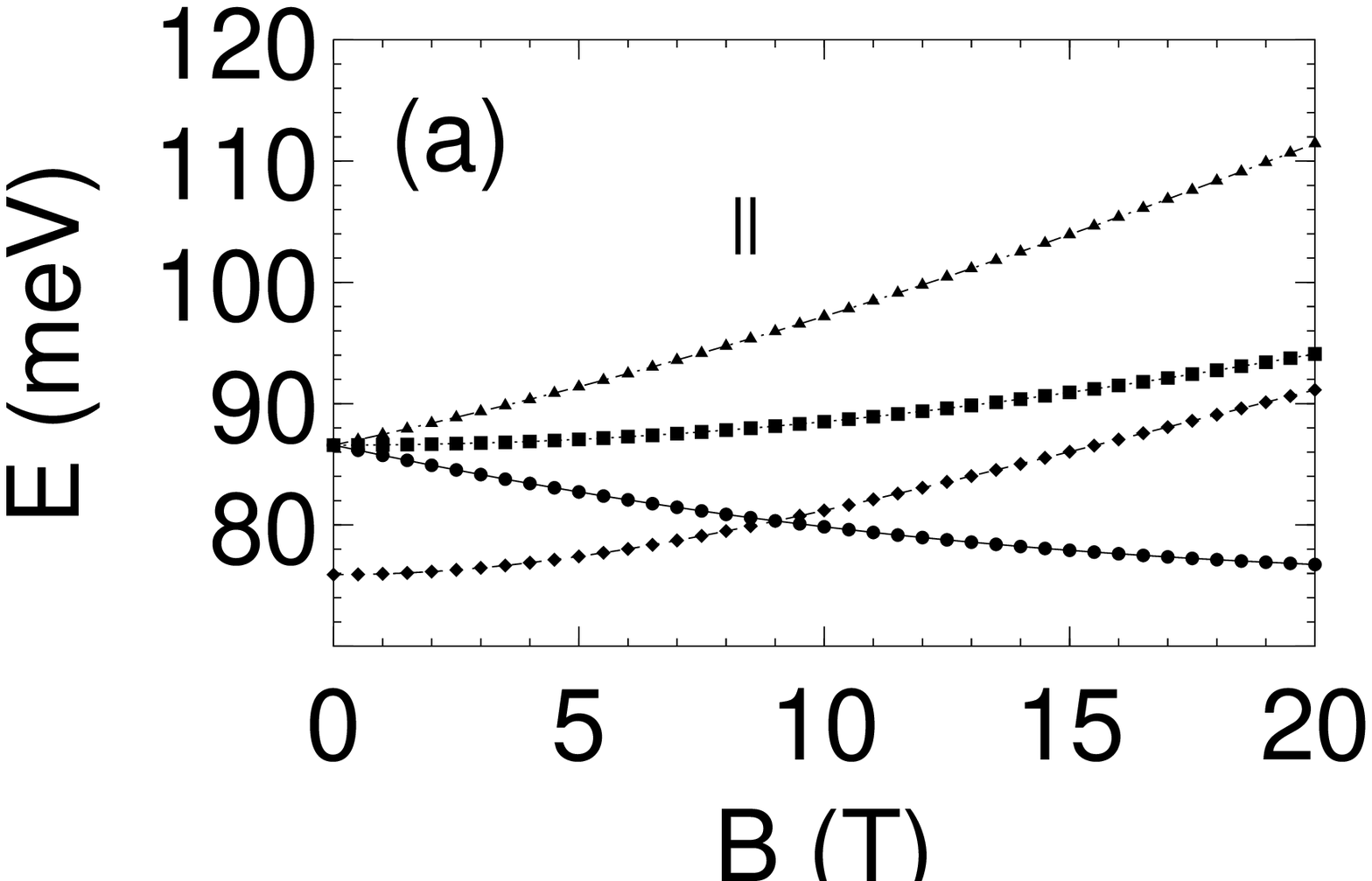,scale=0.24} &
 &
\psfig{file=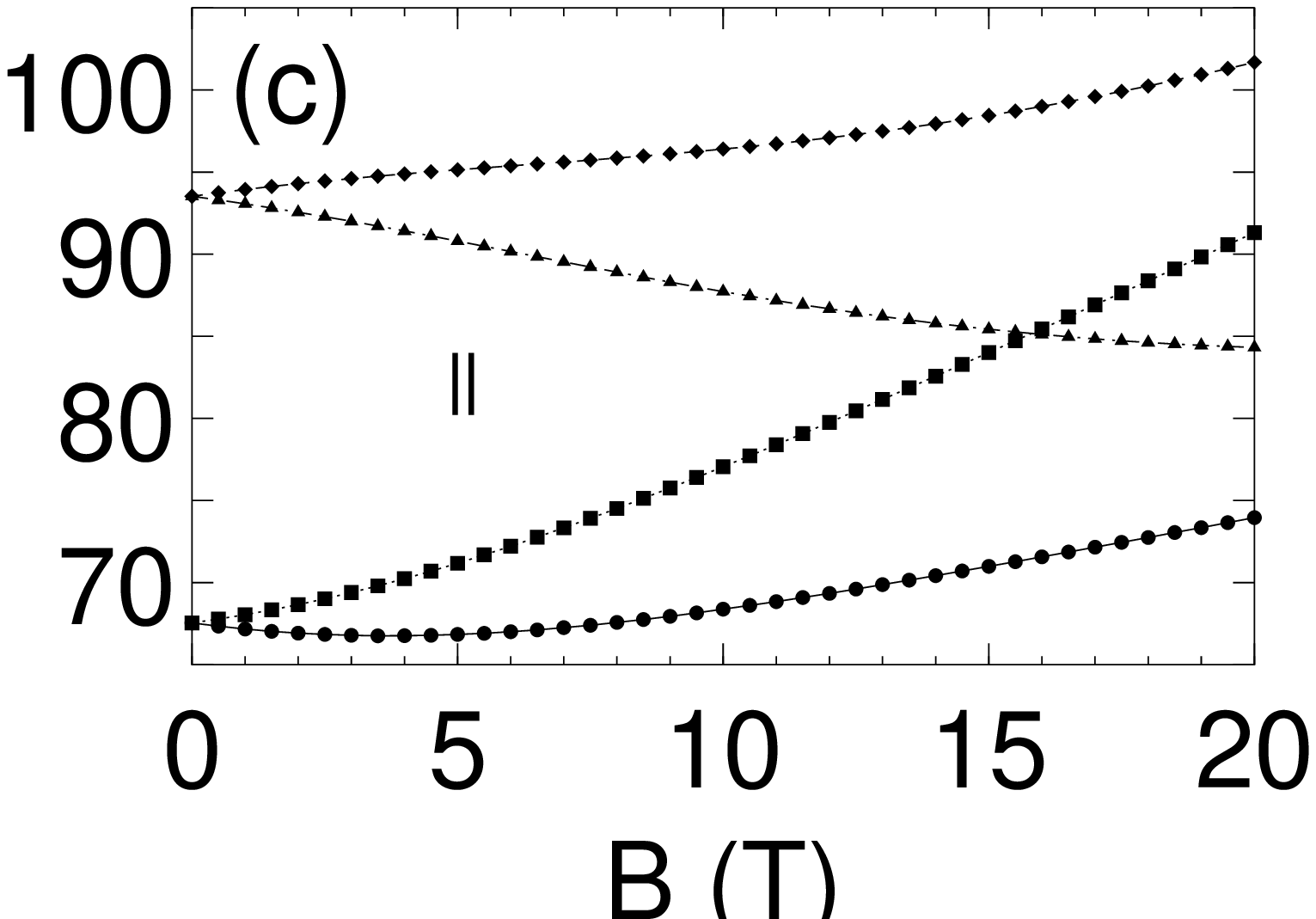,scale=0.24}\\
\hspace{-4mm}
\psfig{file=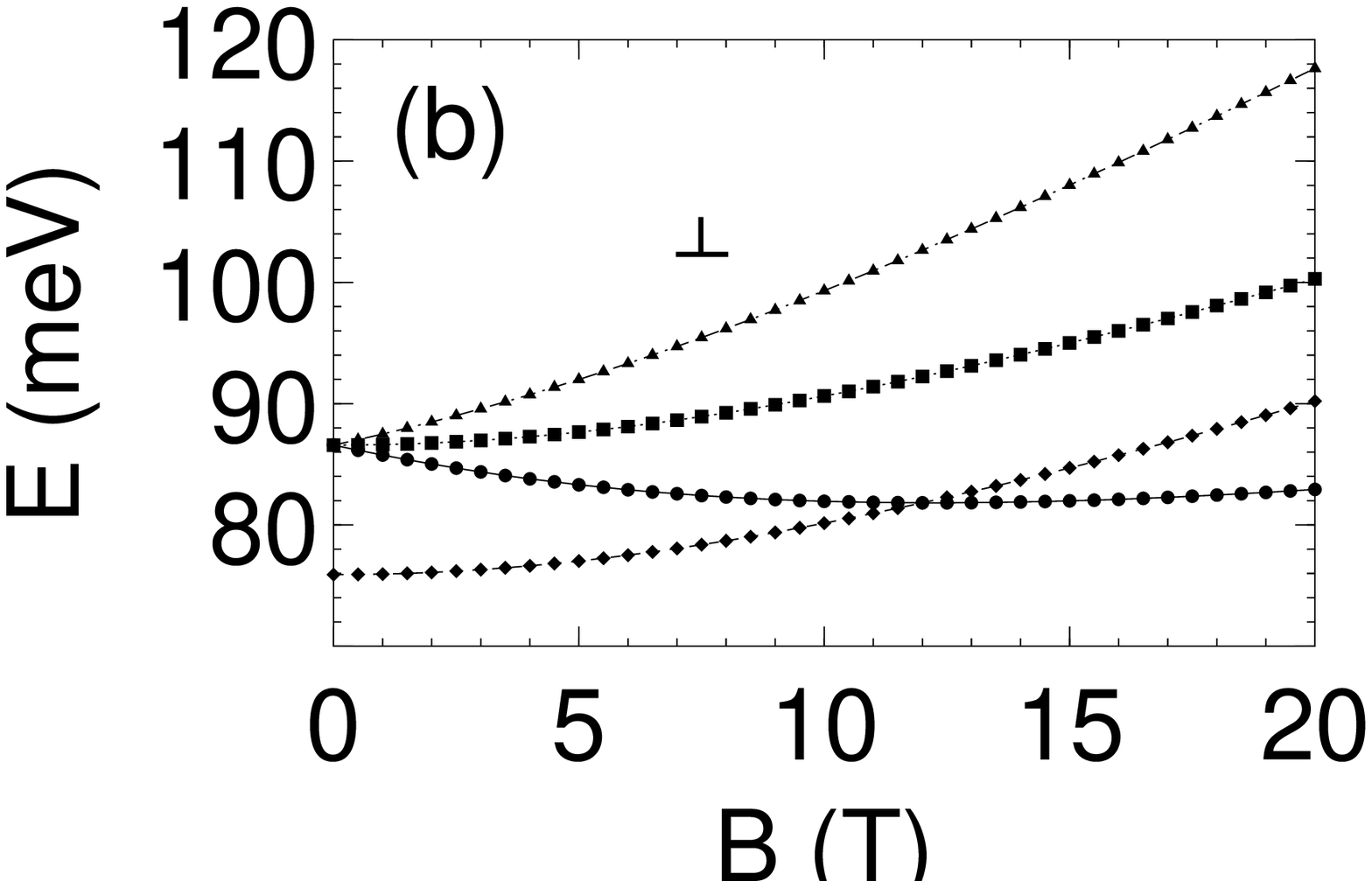,scale=0.24} &
 &
\psfig{file=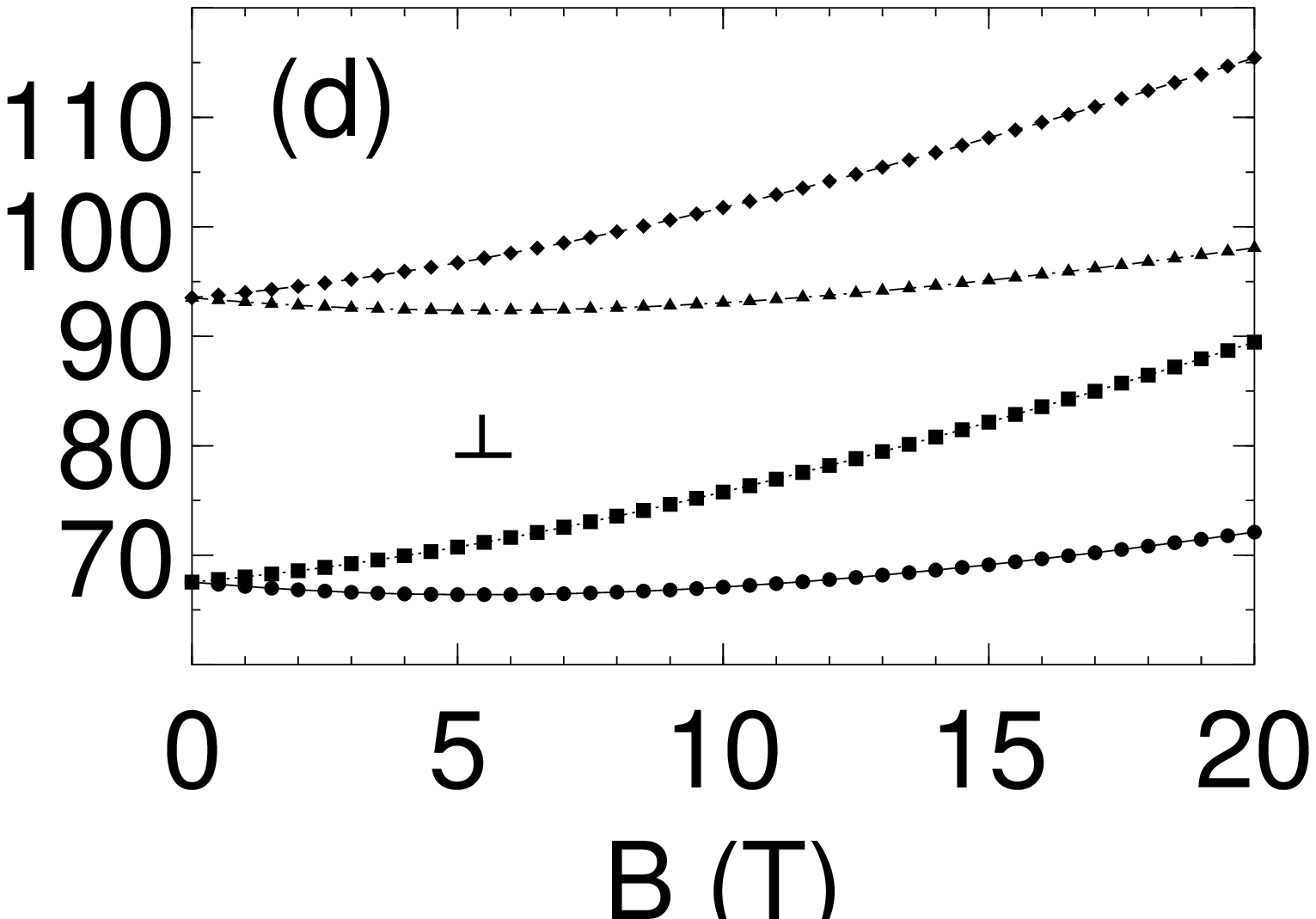,scale=0.24}
\end{tabular}
\caption{\label{osc_fig_2}
Field dependence of the lowest four electronic levels for two
vertically coupled InAs dots (parameters as in
Fig.~\ref{osc_fig_1b}), including the Zeeman coupling
with g-factor $g_{\rm InAs}=-15$.
Left graphs (a,b): Spectrum for a two-electron system involving
the Zeeman-split spin-triplet states (box, circle, and triangle
symbols), and the spin-singlet (diamond symbols). The
exchange energy $J$ corresponds to the gap between the
singlet and the middle ($m_z=0$, box-shaped symbols) triplet
energy. Under the influence of an in-plane field
${\bf B}\parallel x$ (a), the ground state
changes from singlet to triplet
at about $9\,{\rm T}$, whereas in a perpendicular field
${\bf B}\perp x$ (b) the
singlet-triplet crossing occurs at a higher field,
about $12.5\,{\rm T}$.
Right graphs (c,d): Single-particle spectra, again plotted as a
function of $B_\parallel$ (c) and
$B_\perp$ (d). Note that single-particle
and two-particle spectra are clearly distinguishable. In
particular, there is no ground state crossing for a single
electron. The $B$ field dependence of the spectrum of the
large GaAs dots (cf. Fig.~\ref{osc_fig_1}) is similar,
with a much smaller Zeeman splitting
($g_{\rm GaAs}=-0.44$).
The plots are reliable up to a field $B_0\approx 15\,{\rm T}$ where
higher levels start to become important.
}
\end{figure}
We now explain to what extent the Hund-Mulliken results (which we use for
our quantitative evaluations of $J$) are more accurate than the results
obtained from the Heitler-London method (which are more simple and which
we used mostly for qualitative arguments). The Hund-Mulliken method
improves on the Heitler-London method by taking into account double
electron occupancy of the quantum dots.
The Hubbard ratio $t_H/U_H$ can 
be considered a measure for the relative importance of double occupancy.  
Increasing the confinement $\hbar\omega_z$ at constant $d$ (leading to 
potential wells that are deeper but closer together, since
$a=da_B=d\sqrt{\hbar/m\omega_z}$), we observe 
an increase in the discrepancy between $J_{\rm HM}$ and $J_{\rm HL}$ at zero 
magnetic field.  Because the tunneling matrix element $t$ is proportional to 
$\hbar\omega_z$ and the on-site repulsion $U$ is proportional to the Coulomb 
energy $e^2/\kappa a_B\propto \sqrt{\hbar\omega_z}$, the 
Hubbard-ratio $t_H/U_H$ increases as $\sqrt{\hbar\omega_z}$ if the confinement
is increased at constant distance; thus double occupancy becomes more 
important,
explaining the increasing difference between $J_{\rm HM}$ and $J_{\rm HL}$.
Both increasing the inter-dot distance $2a$ and the confinement $\hbar\omega_z$
lead to a larger value of $d=a/a_B$ and thus to a higher tunneling barrier.
A strong decrease of the exchange energy $J$ with increasing $d$ is observed
in both the result calculated according to the Heitler-London and the
Hund-Mulliken approach.

We now turn to the dependence of the exchange energy $J$ on an electric field
$E_\perp$ applied in parallel to the magnetic field, i.e. perpendicular to the
$xy$ plane. Using the Heitler-London approach we find the result
\begin{equation}\label{electric}
J(B,E_\perp) = J(B,0)+ 
\hbar\omega_z\frac{2S^2}{1-S^4}\frac{3}{2}\left(\frac{E_\perp}{E_0}\right)^2,
\end{equation}
where $E_0=m\omega^2_z/e a_B$.  
The growth of $J$ is thus proportional to the square of the electric 
field $E_\perp$, if the field is not too large (see below). 
This result is supported by a Hund-Mulliken calculation, yielding 
the same field dependence at small electric fields, whereas if $eE_\perp a$ is 
larger than $U_H$, double occupancy must be taken into account.
The electric field causes the exchange $J$ at a constant magnetic field $B$
to cross through zero from $J(E=0,B)<0$ to $J>0$.
This effect is signalled by a change in the magnetization $M$,
see Fig.~\ref{osc_fig_4}.

In the presence of an electric field $E_\perp$, the ground-state energy of
an electron in the dot at $z=\pm a$ is 
$\epsilon_{\pm}(E,B)=\hbar\omega_z\left(1+2\alpha_{\pm}(B)-(E/E_0)^2\pm 
2dE/E_0\right)/2$.  The shift of the ground-state energies for 
the upper ($\epsilon_+$) and lower ($\epsilon_-$) dot due to an electric field 
can be used to align the ground-state energy levels of two dots of 
different size (only for two dots of equal size, the energy levels are 
aligned at zero field). This is important because level alignment is
necessary for coherent tunneling 
and thus for the existence of the two-particle singlet and triplet states. 
The parameter $E_a$ denotes the electric field at which the one-particle
ground-states are aligned, $\epsilon_+(B,E_a)=\epsilon_-(B,E_a)$
(for dots of equal size, $E_a=0$).
Investigating the dependence of $J$ on $E_\perp$, one has to be aware
of the fact that
coherent tunneling is suppressed as the electric field is increased,
since the single-particle levels are 
detuned (note, however, that the suppression is not exponential).
This level detuning limits the range of application of 
Eq.~(\ref{electric}), which is only valid for small level misalignment,
$2e(E_\perp-E_a)a<J(0,0)$, where $J(0,0)$ is the exchange
at zero field.  Assuming gates at $20\,{\rm nm}$ below the lower and at 
$20\, {\rm nm}$ above the upper dot in the system discussed above 
($2a\approx 31\,{\rm nm}$, $\hbar\omega_z=16 \,{\rm meV}$ and 
$\alpha_{0}=1/2$), we find that $2aE_\perp e=J(0,0)\approx 0.7\,{\rm meV}$
at a gate voltage of about $U\approx 1.6\,{\rm mV}$.
A further condition for the validity of Eq.~(\ref{electric}) is 
$J(E_\perp)<\hbar\omega_z\alpha_{0-}$, $(\alpha_{0-}\le\alpha_{0+})$.
If this condition is not satisfied, we have to use hybridized single-particle
orbitals.  For the parameters mentioned above, we find
$J(E_\perp)=\hbar\omega_z\alpha_{0-}=8\,{\rm meV}$ at a gate voltage
$U\approx 270\,{\rm mV}$, therefore this condition is automatically fulfilled
if $2eE_\perp a<J(0,0)$.  The numbers used here are arbitrary 
but quite representative, as typical exchange energies are on the order of 
a few meV and inter-dot distances usually range from a few nm to a few
tens of nm.

\begin{figure}
\centerline{\psfig{file=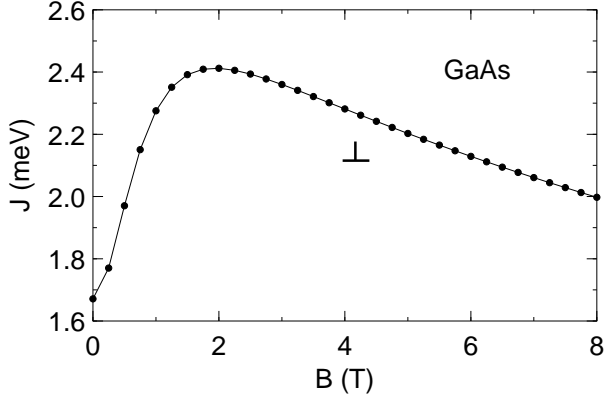,width=8cm}}
\vspace{3mm}
\caption{\label{osc_fig_diff}
Exchange energy $J$ as a function of the perpendicular magnetic
field $B_\parallel$ for two vertically coupled GaAs quantum
dots of different size (both $25\,{\rm nm}$ high, the upper dot
$50\,{\rm nm}$, the lower $100\,{\rm nm}$ in diameter,
$B_{0+}\approx 2\,{\rm T}$).
Here, $J$ is obtained using the Heitler-London method,
Eq.~(\ref{JHLBz}).
The non-monotonic behavior is due to the increase in the overlap,
Eq.~(\ref{overlaposc}), when the orbitals are magnetically 
compressed and therefore the size difference becomes smaller.
}
\end{figure}

In the case where one of the coupled quantum dots is larger than the
other, there is a peculiar non-monotonic behavior when a perpendicular
field $B_\perp$ is applied at $E=0$, see Fig.~\ref{osc_fig_diff}. The
wave-function compression due to the applied magnetic field has the
effect of decreasing the size difference of the two dots, thus
making the overlap Eq.~(\ref{overlaposc}) larger. This growth of
the overlap saturates when the electron orbit in the larger dot
has shrunk approximately to the size of the orbital of the smaller
dot, which happens at roughly $B_{0+}=2mc\omega_z\alpha_{0+}/e$ (assuming
that $\alpha_{0+}\ge\alpha_{0-}$).

\section{In-plane magnetic field $B_\parallel$}\label{parallel}

In this section we consider two dots of equal size in a magnetic field 
$B_\parallel$ which is applied along the $x$-axis, i.e. \textit{in-plane}
(see Fig.~\ref{potential}). Since the two dots have the same 
size, the lateral confining potential Eq.~(\ref{lateral}) reduces to 
$V(x,y)=m\omega^2_{z}\alpha^2_0(x^2+y^2)/2$, where the parameter $\alpha_0$
describes the ratio between the lateral and the vertical confinement energy.
The vertical double-dot structure is modeled using the potential
Eq.~(\ref{VOsc}). The single-dot Hamiltonian is given by Eq.~(\ref{h0})
with the vector potential ${\bf A}({\bf r})=B(0,-z,y)/2$. The
situation for an in-plane field is a bit more complicated than for
a perpendicular field, because the planar and vertical motion do
not separate.
In order to find the ground-state wave function of the one-particle
Hamiltonian $h^0_{\pm a}$, we have applied the variational method
(cf.  Appendix \ref{appendixBx}), with the result
\begin{eqnarray}
\varphi_{\pm a}({\bf r}) &=&
\left(\frac{m \omega_z}{\pi\hbar}\right)^{3/4}\left(\alpha_{0} \alpha\beta 
\right)^{1/4}\exp\left(\pm i\frac{ya}{2l_B^2}\right)\label{var}\\
&&\times
\exp\left[-\frac{m\omega_z}{2\hbar} \left(\alpha_{0} x^2+\alpha 
y^2+\beta(z \mp a)^2\right)\right].\nonumber
\end{eqnarray}
Note that this is not the exact single-dot groundstate, except
for spherical dots ($\alpha_0= 1$).
We have introduced the parameters $\alpha(B)=\sqrt{\alpha^2_0+(B/B_0)^2}$ and 
$\beta(B)=\sqrt{1+(B/B_0)^2}$ describing the wave-function compression in 
$y$ and $z$ direction, respectively.  The phase factor involving the 
magnetic length $l_B=\sqrt{\hbar c/eB}$ is due to the gauge transformation 
${\bf A}_{\pm a}=B(0,-(z\mp a),y)/2\rightarrow {\bf A}=B(0,-z,y)/2$.
The one-particle ground-state energy amounts to
$\epsilon_0=\hbar\omega_z\left(\alpha_{0}+\alpha+\beta\right)/2$.
From $\varphi_{\pm a}$ we  construct a symmetric and an antisymmetric
two-particle wave function $\Psi_\pm$, exactly as for ${\bf B}\parallel z$.
Care has to be taken calculating the exchange energy $J$;
Eq.~(\ref{Jformal}) has to be modified, since $\varphi_{\pm a}$ is not 
an exact eigenstate of the Hamiltonian $h^0_{\pm a}$ (cf.  Appendix 
\ref{appendixBx}).
The correct expression for the exchange energy in this case 
is \begin{equation}\label{JformalBx} 
J(B,d)=J_0(B,d)
-\hbar\omega_z\frac{4S^2}{1-S^4}\frac{\beta 
-\alpha}{\alpha}d^2\left(\frac{B}{B_0}\right)^2,
\end{equation} 
where $J_0$ denotes the expression from Eq.~(\ref{Jformal}).
The variation of the exchange energy $J$ as a function of the
magnetic field $B$ is, through the prefactor $2S^2/(1-S^4)$,
determined by the overlap
\begin{equation} 
S(B,d)=\exp\left[-d^2\left(\beta (B)+\frac{1}{\alpha (B)}
\left(\frac{B}{B_0}\right)^2\right)\right],
\end{equation}
depending exponentially on the in-plane field, while for a perpendicular
field the overlap is independent of the field (for two dots of 
equal size), see Eq.~(\ref{overlaposc}).
We find that for weakly confined dots ($\hbar\omega_z\lesssim 10\,{\rm meV}$),
there is a singlet-triplet crossing ($J$ becoming negative), e.g. for
$\hbar\omega_z=7\,{\rm meV}$, $\alpha_0=1/2$, and $2a=25\,{\rm nm}$
we find a singlet-triplet crossing  at $B\approx 6\,{\rm T}$.
No singlet-triplet crossing is found for systems with vertical confinement 
$\hbar\omega_z\gtrsim 10\,{\rm meV}$.
Generally, the decay of $J$ becomes flatter as the confinement is increased.
Improving on the Heitler-London result, we have again performed
a molecular-orbital
(Hund-Mulliken) calculation of the exchange energy, which
we plot in Fig.~\ref{osc_fig_1} (left graph, circle symbols).
\begin{figure}
\begin{tabular}{r r r}
\hspace{-4mm}
\psfig{file=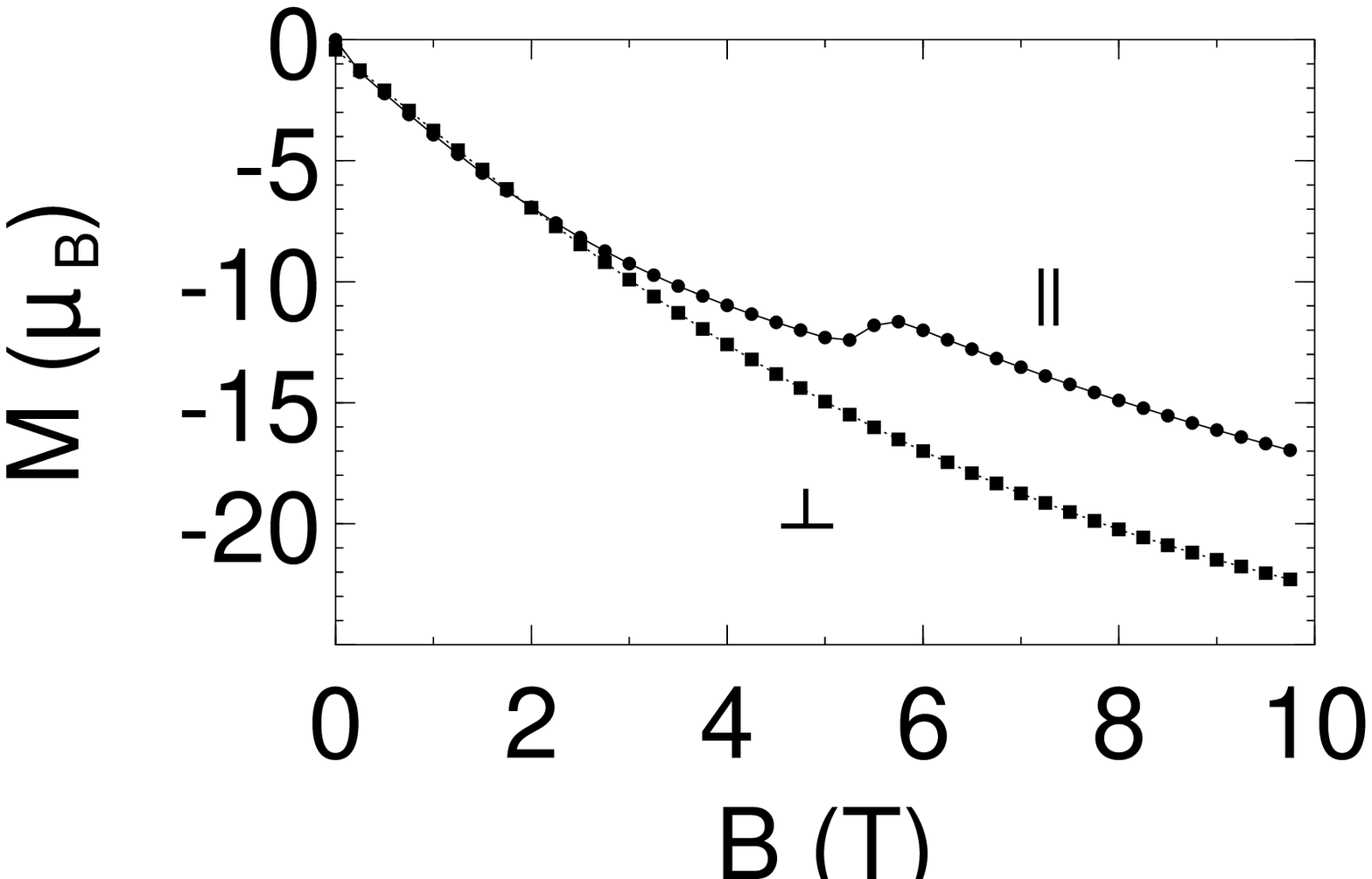,scale=0.24} &
 &
\psfig{file=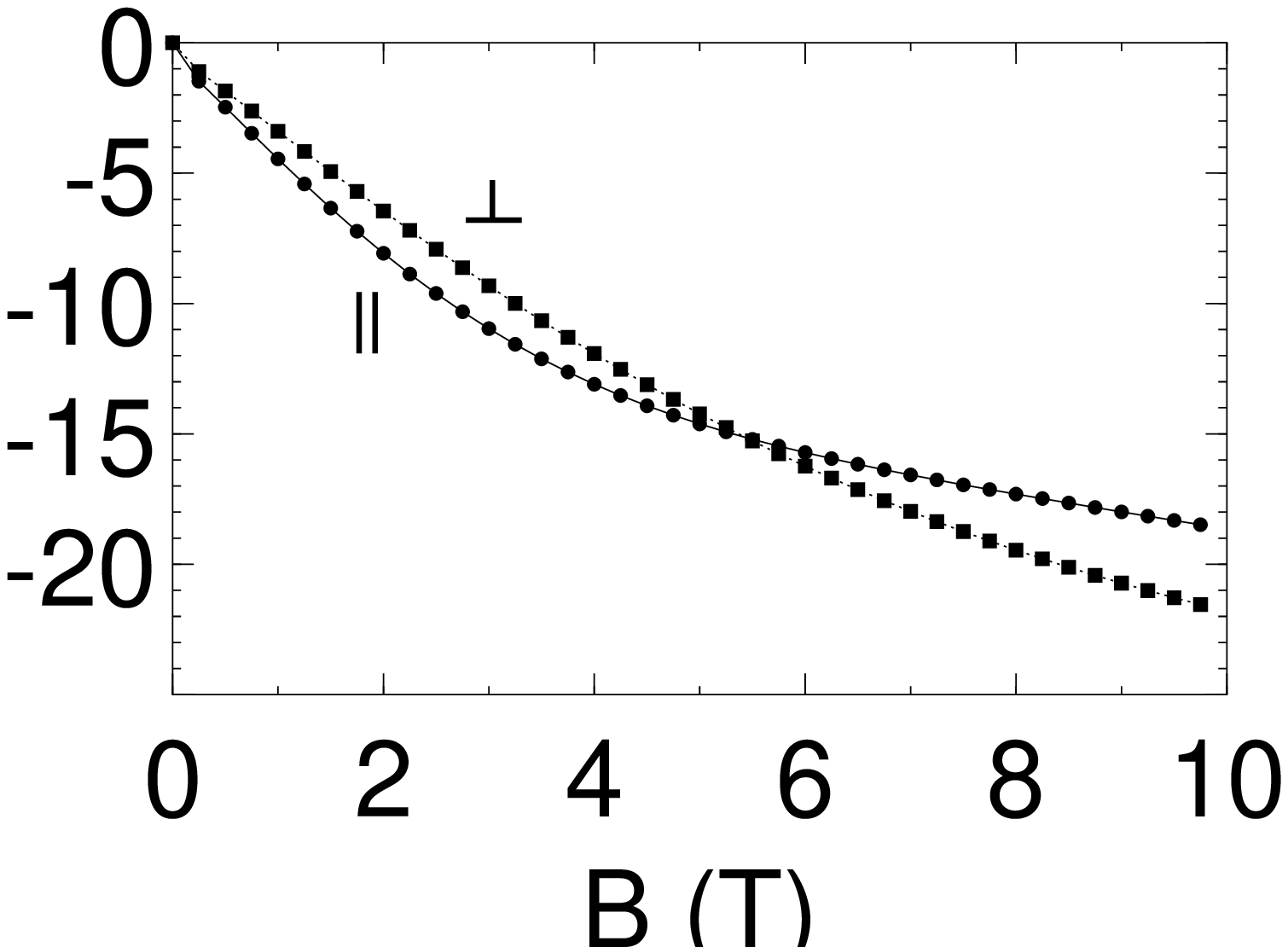,scale=0.24}
\end{tabular}
\caption{\label{osc_fig_3}
Magnetization $M$ (in units of Bohr magnetons) as a function of
the $B$ field for vertically coupled large GaAs ($g=-0.44$)
quantum dots (parameters as in Fig.~\ref{osc_fig_1}) containing two
electrons (left graph) and a single electron only (right graph)
at $T=100\,{\rm mK}$.
The box shaped symbols correspond to $B_\perp$, the circles to
$B_\parallel$. The singlet-triplet crossing in the two-electron 
system (due to the Zeeman splitting and the decrease of $J$) causes
a jump in the magnetization around $5.5\,{\rm T}$
for $B_\parallel$, but no such signature occurs for $B_\perp$.
}
\end{figure}

\begin{figure}
\begin{tabular}{r r r}
\hspace{-4mm}
\psfig{file=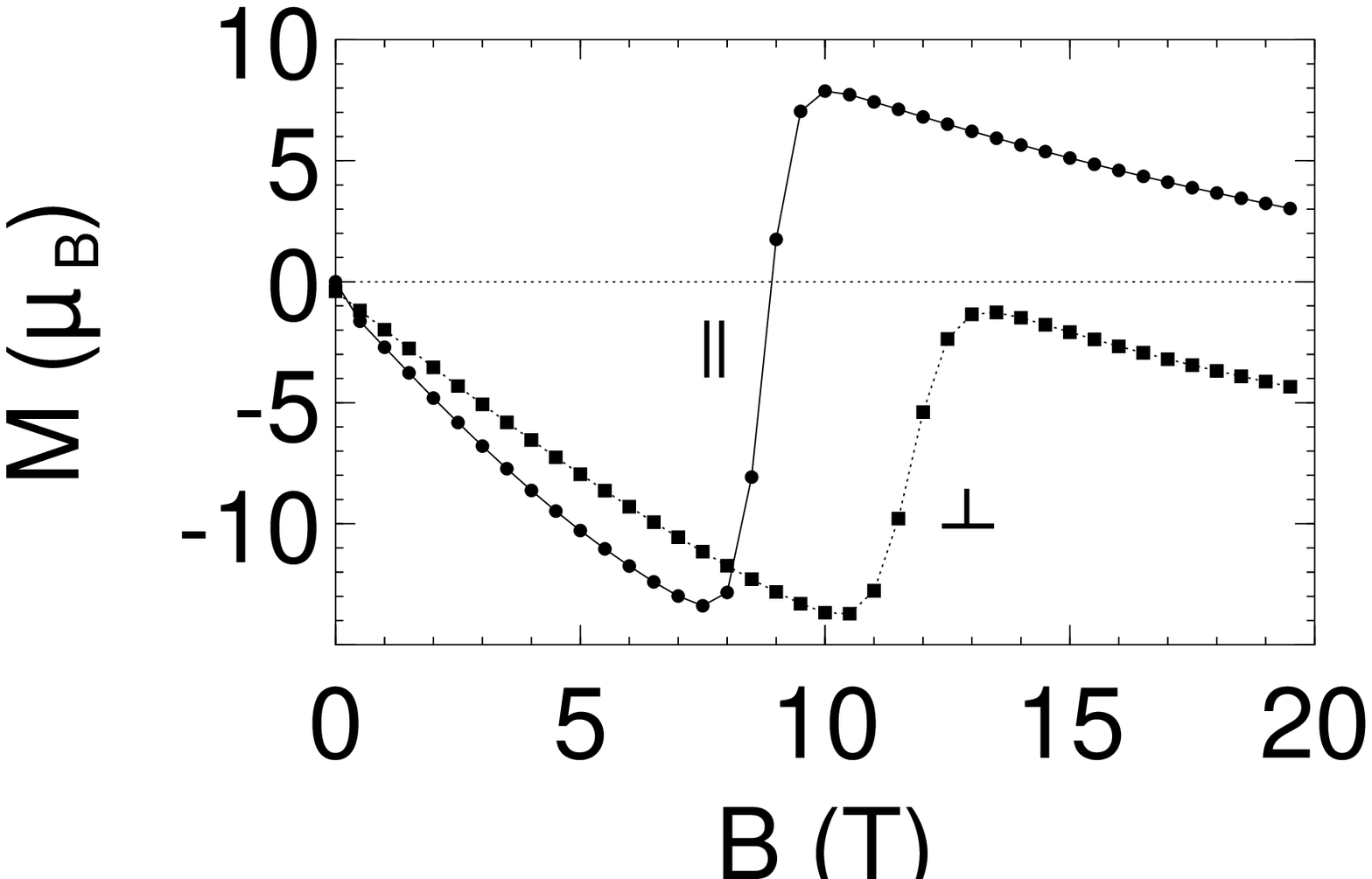,scale=0.24} &
 &
\psfig{file=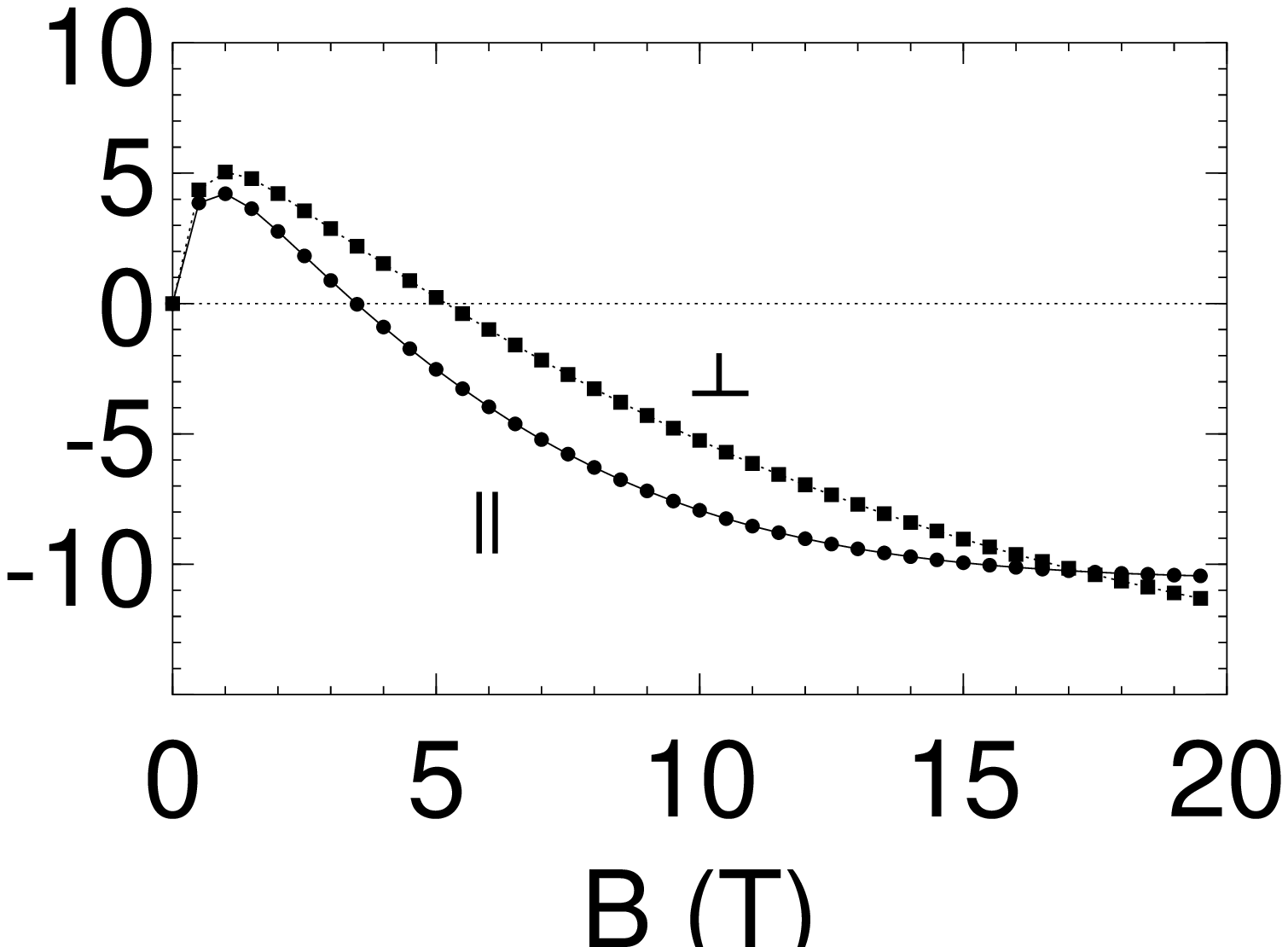,scale=0.24}
\end{tabular}
\caption{\label{osc_fig_3b}
Magnetization $M$ (in units of Bohr magnetons) as a function of
the $B$ field for vertically coupled small InAs ($g=-15$)
quantum dots (parameters as in Fig.~\ref{osc_fig_2}) containing two
electrons (left graph) and a single electron only (right graph)
at $T=4\,{\rm K}$.
The box-shaped symbols correspond to $B_\perp$, the circles to
$B_\parallel$. The singlet-triplet crossing in the two-electron 
system causes a jump in the magnetization around $9\,{\rm T}$
for $B_\parallel$, and at about $12.5\,{\rm T}$ for $B_\perp$.
}
\end{figure}

\begin{figure}
\begin{tabular}{r r r}
\hspace{-4mm}
\psfig{file=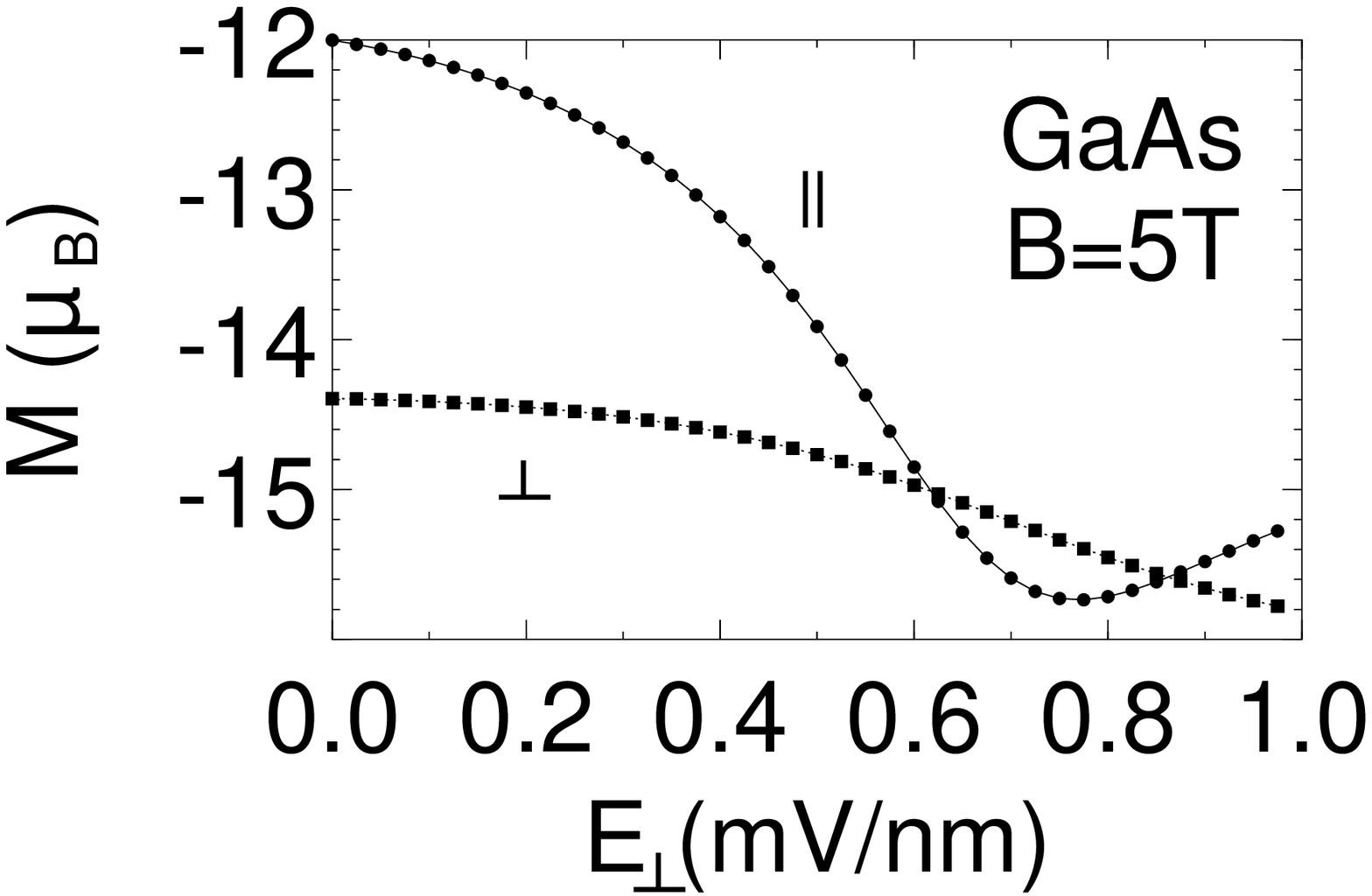,scale=0.24} &
 &
\psfig{file=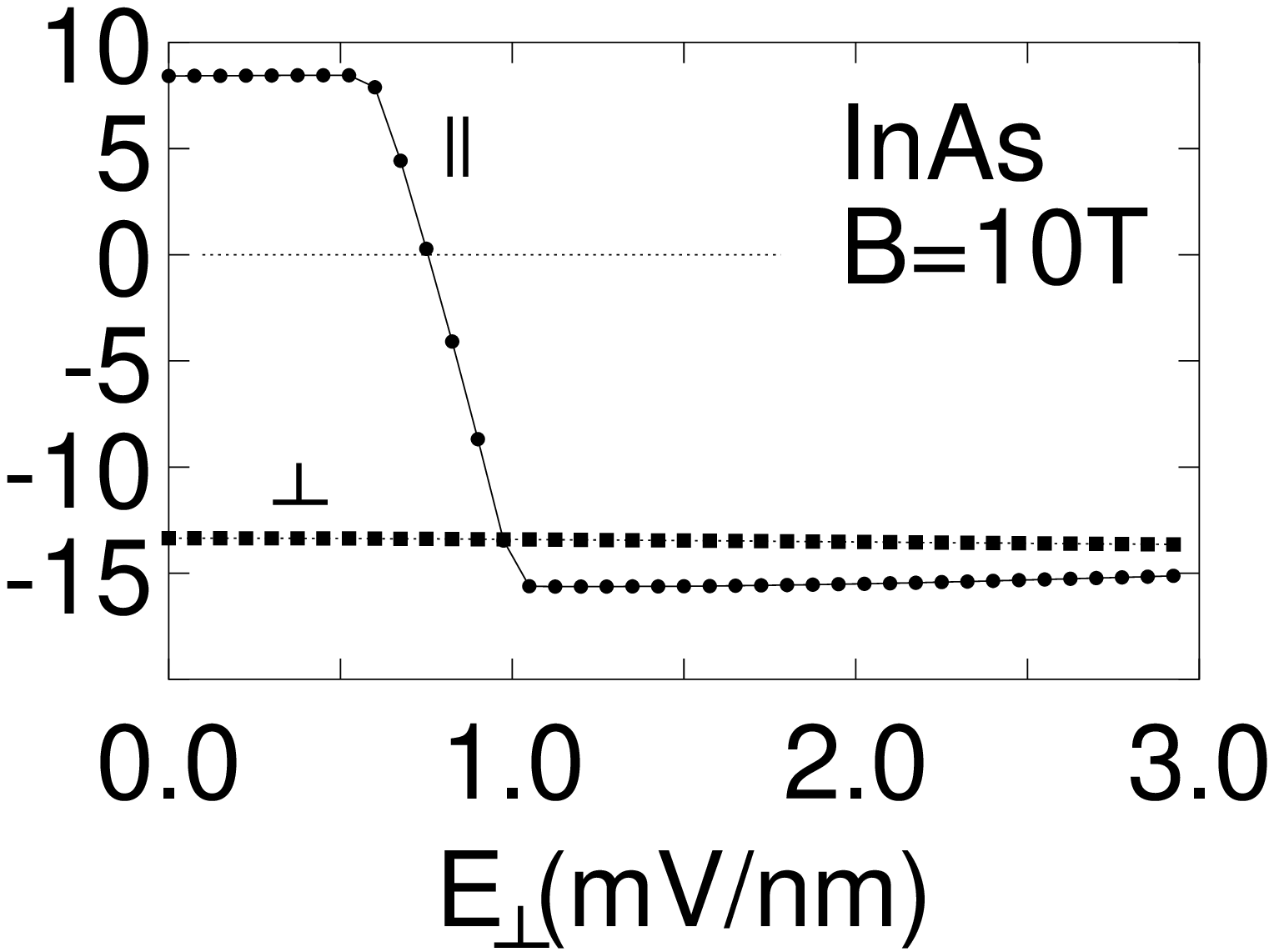,scale=0.24}
\end{tabular}
\caption{\label{osc_fig_4}
Magnetization $M$ (in units of Bohr magnetons) as a function of
the perpendicular electric field $E_\perp$ for vertically coupled
quantum dots containing two electrons at fixed magnetic field.
The box-shaped symbols correspond to $B_\perp$, the circles to
$B_\parallel$. Starting at $E=0$ with a triplet groundstate for $B_\parallel$
(not so for $B_\perp$), the electric field eventually causes
a change of the groundstate back to the singlet, which leads
again to a jump in the magnetization for $B_\parallel$.
The left graph corresponds to a GaAs
double-dot (parameters as in Fig.~\ref{osc_fig_1}) at $T=100\,{\rm mK}$
and $B=5\,{\rm T}$,
whereas the right graph is for a smaller InAs double-dot 
(as in Fig.~\ref{osc_fig_1b})
at $T=4\,{\rm K}$ and $B=10\,{\rm T}$.
}
\end{figure}

It is crucial in experiments to distinguish between single-
and two-electron effects in the double dot.
A single electron in a double dot exhibits a level splitting
of $2t$, where $t$ denotes the single-particle 
tunneling matrix element, cf. Eq.~(\ref{t}), which has a
$B$ field dependence similar to the exchange coupling $J$.
In order to allow a distinction between $J$ and $t$, we have plotted
$t(B)$ in the right graph of Fig.~\ref{osc_fig_1}.
Since the one-particle tunneling matrix 
element $t$ is strictly positive, it is clearly distinguishable from the 
exchange energy $J$ in systems with singlet-triplet crossing.
Experimentally, the number of electrons in the double-dot system can
be tested via the field-dependent spectrum (Fig.~\ref{osc_fig_2})
and magnetization (Figs.~\ref{osc_fig_3}-\ref{osc_fig_4}).

\section{Electrical switching of the spin interaction}\label{e_switch}

\begin{figure}
\centerline{\psfig{file=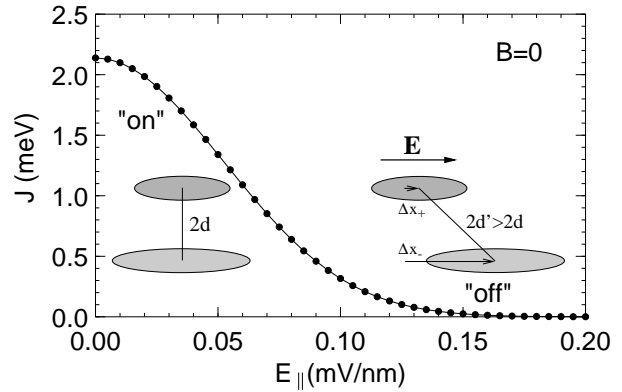,width=8cm}}
\vspace{5mm}
\caption{\label{switch}
Switching of the spin-spin coupling between dots of different 
size by means of an in-plane electric field $E_\parallel$ ($B=0$).
The exchange coupling is switched ``on'' at $E=0$. When
an in-plane electric field $E_\parallel$ is applied, the larger
of the two dots is shifted to the right by $\Delta x_-$,
whereas the smaller dot is shifted by $\Delta x_+<\Delta x_-$,
where $\Delta x_\pm =E_\parallel/E_0\alpha_{0\pm}^2$ and
$E_0=\hbar\omega_z/ea_B$.
Therefore, the mean distance between the
electrons in the two dots grows as
$d'=\sqrt{d^2+A^2(E_\parallel/E_0)^2}$, where
$A=(\alpha_{0+}^2-\alpha_{0-}^2)/2\alpha_{0+}^2\alpha_{0-}^2$.
The exchange coupling $J$, being exponentially sensitive to the
inter-dot distance $d'$, decreases thus exponentially,
$J\approx S^2\approx \exp[-2A^2(E_\parallel/E_0)^2]$.
We have chosen $\hbar\omega_z=7\,{\rm meV}$, $d=1$,
$\alpha_{0+}=1/2$ and $\alpha_{0-}=1/4$.
For these parameters, we find
$E_0=\hbar\omega_z/ea_B=0.56\,{\rm mV/nm}$ and
$A=(\alpha_{0+}^2-\alpha_{0-}^2)/2\alpha_{0+}^2\alpha_{0-}^2=6$.
The exchange coupling $J$ decreases exponentially on the scale
$E_0/2A = 47 \,{\rm mV/\mu m}$ for the electric field.
}
\end{figure}

Coupled quantum dots can potentially be used as quantum gates for
quantum computation\cite{loss,burkard}, where the electronic spin
on the dot plays the role of the qubit. Operating a coupled quantum
dot as a quantum gate requires the ability to switch on and off
the interaction between the electron spins on neighboring dots.
We present here a simple method of achieving a high-sensitivity
switch for vertically coupled dots by means of a horizontally
applied electric field $E_\parallel$.
The idea is to use a pair of quantum dots with different lateral
sizes, e.g. a small dot on top of a large dot
($\alpha_{0+}>\alpha_{0-}$, see Fig.~\ref{potential}).
Note that only the radius in the $xy$ plane has to be different,
while we assume that the dots have the same height.
Applying an in-plane electric field $E_\parallel$ in this case
causes a shift of the single-dot orbitals by
$\Delta x_\pm = eE_\parallel/m\omega_z^2\alpha_{0\pm}^2
=E_\parallel/E_0\alpha_{0\pm}^2$, where $E_0=\hbar\omega_z/ea_B$, see
Fig.~\ref{switch}.
It is clear that the electron in the larger dot moves further in the
(reversed) direction of the electric field ($\Delta x_- >\Delta x_+$),
since its confinement potential is weaker. As a result, the mean
distance between the two electrons changes from $2d$ to $2d'$, where
\begin{equation}
  \label{d_electric}
  d' = \sqrt{d^2 + \frac{1}{4}\left(\Delta x_- - \Delta x_+\right)^2}
     = \sqrt{d^2 + A^2\left(\frac{E_\parallel}{E_0}\right) ^2},
\end{equation}
with $A=(1/\alpha_{0-}^2-1/\alpha_{0+}^2)/2$.
Using Eq.~(\ref{overlaposc}), we
find that $S\propto \exp(-d'^2)\propto \exp[-A^2(E_\parallel/E_0)^2]$,
i.e. the orbital overlap decreases exponentially as a function
of the applied electric field $E_\parallel$. Due to this high sensitivity,
the electric field is an ideal ``switch'' for the exchange coupling
$J$ which is (asymptotically) proportional to $S^2$ and thus decreases
exponentially on the scale $E_0/2A$. Note that if
the dots have exactly the same size, then $A=0$ and the effect vanishes.
We can obtain an estimate of $J$ as a function of $E_\parallel$
by substituting $d'$ from Eq.~(\ref{d_electric}) into the
Heitler-London result, Eq.~(\ref{JHLBz}). A plot of $J(E_\parallel)$
obtained in this way is shown in Fig.~\ref{switch} for a
specific choice of GaAs dots. Note that this procedure is not
exact, since it neglects the tilt of the orbitals with respect to their
connecting line. 
Exponential switching is highly desirable for quantum computation,
because in the ``off'' state of the switch, fluctuations in the
external control parameter (e.g. the electric field $E_\parallel$)
or charge fluctuations
cause only exponentially small fluctuations in the coupling $J$.
If this were not the case, the fluctuations in $J$ would lead to
uncontrolled coupling between qubits and therefore to multiple-qubit
errors. Such correlated errors cannot be corrected by known
error-correction schemes, which are designed for
uncorrelated errors\cite{preskill}.
It seems likely that our proposed switching method can be realized
experimentally, e.g. in vertical columnar GaAs quantum dots\cite{austing}
with side gates controlling the lateral size and position of the dots,
or in SADs where one can expect different dot sizes anyway.

\section{Spin measurements}\label{spin_meas}

The magnetization (Figs.~\ref{osc_fig_3}-\ref{osc_fig_4}),
measured as an ensemble average over many pairs of coupled quantum dots in
thermal equilibrium, reveals whether the ground-state of the coupled-dot
system is a spin singlet or triplet. On the one hand, such a magnetization
could be detected by a
SQUID or with cantilever-based\cite{wago,harris} magnetometers.
This type of spin measurement was already suggested earlier for laterally
coupled dots\cite{burkard}.
The distinction between spin singlet ($S=0$) and triplet ($S=1$) is also
possible using optical methods: Measurement of the Faraday rotation
(caused by the precession of the magnetic moment around a magnetic field)
using
a pump-probe technique\cite{kikkawa,gupta} reveals if the two-electron
system is in a singlet ($S=0$) state with no Faraday rotation or in a
triplet ($S=1$) state with finite Faraday rotation. 
Finally, it should also be possible to obtain spin information via
optical (far-infrared) spectroscopy\cite{jacak}.

We remark that if it is possible to measure the magnetization
of just one individual pair of coupled dots, then this is equivalent to
measuring a microscopic two spin-1/2 system, i.e. $1/2 \otimes 1/2=0\oplus
1$. We have described elsewhere
how such individual singlet and triplet states in a double dot can be
detected (through their charge) in transport  measurements
via Aharonov-Bohm oscillations in the
cotunneling current and/or current correlations\cite{MMM2000,BLS,LS}.

It is interesting to note that above scheme allows one to 
measure even a single spin $1/2$, provided that, in addition,
one can perform one two-qubit gate operation (corresponding to switching
on the coupling $J$ for some
finite time) and a subsequent single-qubit gate by means of applying a
local Zeeman interaction to one of the qubits. [Such local Zeeman
interactions can be generated e.g. by using local magnetic fields or by
inhomogeneous g-factors \cite{MMM2000}.]
Explicitly, such a single-spin measurement of the electron is performed as
follows. We are given an arbitrary spin $1/2$ state $|\alpha\rangle$
in quantum dot 1. For simplicity, we assume that $|\alpha\rangle$ is one
of the basis states, $|\alpha\rangle=|\!\uparrow\rangle$ or
$|\alpha\rangle=|\!\downarrow\rangle$; the generalization to a
superposition of the basis states is straightforward.
The spin in quantum dot 2 is prepared in the state $|\!\uparrow\rangle$.
The interaction J between the spins in Eq. (\ref{Heisenberg})
is then
switched on for a time
$\tau_s$, such that $\int_0^{\tau_s}J(t)dt=\pi/4$. By doing this, a
`square-root-of-swap' gate\cite{loss,burkard2} is performed for the two
spins (qubits).
In the case $|\alpha\rangle=|\!\uparrow\rangle$,
nothing happens, i.e. the spins remain in the state
$|\!\uparrow\uparrow\rangle$, whereas if
$|\alpha\rangle=|\!\downarrow\rangle$, then we obtain the
entangled state
$(|\!\downarrow\uparrow\rangle +i|\!\uparrow\downarrow\rangle)/\sqrt{2}$,
(up to a phase factor which can be ignored). Finally, we apply
a local Zeeman term, $g\mu_B B S_z^1$,
acting parallel to the $z$-axis at quantum dot 1
during the time interval $\tau_B$, such that $\int_0^{\tau_B}(g\mu_B
B) (t)dt =
\pi/2$. The resulting state is (again up to unimportant phase factors)
the triplet state $|\!\uparrow\uparrow\rangle$ in the case where
$|\alpha\rangle=|\!\uparrow\rangle$, whereas we obtain the singlet state
$(|\!\uparrow\downarrow\rangle - |\!\downarrow\uparrow\rangle)/\sqrt{2}$
in the case $|\alpha\rangle=|\!\downarrow\rangle$. 
In other words, such a procedure maps the triplet
$|\!\uparrow\uparrow\rangle$ into itself and the state
$|\!\downarrow\uparrow\rangle$ into the singlet (similarly, 
the same gate operations map
$|\!\downarrow\downarrow\rangle$ into itself, while
$|\!\uparrow\downarrow\rangle$ is mapped into the
triplet $(|\!\uparrow\downarrow\rangle +
|\!\downarrow\uparrow\rangle)/\sqrt{2}$, again up to phase factors).
Finally, measuring 
the total magnetic moment of the double dot  system then reveals
which of the two spin states in dot 1,  $|\!\uparrow\rangle$ or 
$|\!\downarrow\rangle$, was realized initially.

\section{Discussion}

In summary, we have calculated the spin exchange interaction
$J({\bf B},{\bf E})$ for
electrons confined in a pair of vertically coupled quantum dots,
and have compared the two-electron spectra (with level splitting $J$) to
the single-electron spectra (with level splitting $2t$).
Comparing the one- and two-electron spectra enables us to distinguish
one-electron filling from two-electron filling of the double dot in an
experiment. For two-electron filling in the presence of a magnetic field,
a ground-state crossing from singlet to triplet occurs at fields of about
$5$ to $10\,{\rm T}$, depending on the strength of the confinement, the
coupling, and the effective g-factor. The crossing can be reversed by
applying a perpendicular electric field.

As a model for the electron confinement in a quantum dot we have 
chosen harmonic potentials. However, in some situations (especially
self-assembled quantum dots) it is more accurate to use
square-well confinement potentials in order to model the band-gap
offset between different materials. We have also performed
calculations using square-well potentials, which confirm
the qualitative behavior of the results obtained using
harmonic potentials. The results from using the square-well
model potentials cannot be written in simple algebraic expressions,
and are given elsewhere\cite{seelig}.

Furthermore, we have analyzed the possibilities of switching the
spin-spin interaction $J$ using external parameters.
We find that in-plane magnetic fields $B_\parallel$ (perpendicular to the 
inter-dot axis) are better suited for tuning the exchange coupling in a 
vertical double-dot structure than a field $B_\perp$ (applied along the
inter-dot axis).  Moreover, we have confirmed that the dependence of the
exchange energy on a magnetic field is stronger for weakly confined 
dots than for structures with strong confinement.
An even more efficient switching mechanism is found when a small 
quantum dot is coupled to a large dot: In this case, the coupling $J$
depends exponentially on the in-plane electric field $E_\parallel$,
and thus provides an ideal external parameter for switching on and
off the spin coupling with exponential sensitivity.
The experimental confirmation of the
electrical switching effect would be an important step towards
solid-state quantum computation with quantum dots.

Another (very demanding) key experiment for quantum computation in
quantum dots is the measurement of single electron spins.
We have presented here a theoretical scheme for a single-spin
measurement using coupled quantum dots.
Obviously this scheme already requires some controlled interaction
between the spins (qubits) and therefore the successful
implementation of some switching mechanisms.

\acknowledgments
We would like to thank D.D. Awschalom, D.P. DiVincenzo, R.H. Blick, P.M.
Petroff, and E.V. Sukhorukov for useful discussions.
This work is supported by the Swiss National Science Foundation.

\appendix

\section{Hund-Mulliken matrix elements}\label{appendix_matrixelements}
 
Here we list the explicit expressions for the matrix elements
defined in Eqs.~(\ref{matrix})--(\ref{C_offdiag}) for two dots with
arbitrary
(and possibly different) single-electron Hamiltonians $h_{\pm a}$ and
(non-orthogonal) single-electron orbitals $\varphi_{\pm a}$ centered at
$z=\pm a$.
The matrix elements are
\begin{eqnarray}
V_+ &=& N^4\left[2g^2(G_1^++G_1^-)+4g^2S^2G_1^0+4g^2G_2\right.
\nonumber\\
&&\quad\quad\quad\quad
\left.+(1+g^2)^2G_3-6g^2(G_4^+ + G_4^-)\right],\label{HMquant1}\\
V_- &=& N^4(1-g^2)^2\left[G_3-S^2G_2\right],\label{HMquant2}\\
U_\pm &=& N^4\left[G_1^\pm+g^4G_1^\mp +2g^2S^2G_1^0+2g^2S^2G_2
\right.\nonumber\\
&&\quad\quad\quad\quad
\left.+2g^2S^2G_3-4gS(G_4^\pm-g^2G_4^\mp)\right],
\label{HMquant3}\\
X &=& N^4\left[(1+g^4)G_1^0S^2+g^2(G_1^+ + G_1^-)+2g^2S^2G_2
\right.\nonumber\\
&&\left.\quad\quad\quad
+2g^2G_3-2g(1+g^2)S(G_4^+ + G_4^-)\right],
\label{HMquant4}\\
w_\pm &=& N^4\left[-g G_1^\pm-g^3G_1^\mp
-g(1+g^2)(2S^2G_1^0+G_3)\right.\nonumber\\
&&\quad\quad\quad
\left. +S(1+3g^2)G_4^\pm+S^2g^2(1+g^2)G_4^\mp\right],\label{HMquant5}
\end{eqnarray}
with $N=1/\sqrt{1-2Sg+g^2}$ and $g=(1-\sqrt{1-S^2})/S$.
We have introduced the overlap integrals
\begin{eqnarray}
G_1^\pm &=& \langle\varphi_{\pm a}\varphi_{\pm a}|C|\varphi_{\pm a}\varphi_{\pm a}\rangle,\label{integral1}\\
G_1^0 &=& S^{-2} \langle\varphi_{\pm a}\varphi_{\pm a}|C|\varphi_{\mp a}\varphi_{\mp a}\rangle,\label{integral2}\\
G_2 &=& S^{-2} \langle\varphi_{\pm a}\varphi_{\mp a}|C|\varphi_{\mp a}\varphi_{\pm a}\rangle,\label{integral3}\\
G_3 &=& \langle\varphi_{\pm a}\varphi_{\mp a}|C|\varphi_{\pm a}\varphi_{\mp a}\rangle,\label{integral4}\\
G_4^\pm &=& S^{-1} 
\langle\varphi_{\pm a}\varphi_{\pm a}|C|\varphi_{\pm a}\varphi_{\mp a}\rangle.\label{integral5}
\end{eqnarray}
Note that the expressions for $G_1^0$, $G_2$, and $G_3$ are invariant under
exchange of $\varphi_{a}$ and $\varphi_{-a}$.
In the case where the two single-particle Hamiltonians coincide (implying
that the dots have the same size), we find $G_1^+=G_1^-$ ($=G_1^0$, since
$C$ depends only on the relative coordinate) and
$G_4^+=G_4^-$, and the expressions in
Eqs.~(\ref{HMquant1})--(\ref{HMquant5}) for the matrix
elements can be simplified accordingly. This simplification leads
to the same form of the Hund-Mulliken matrix elements which we have
calculated for laterally coupled dots\cite{burkard}.
If it is possible to choose the orbitals $\varphi_{\pm a}$ to be real,
e.g. if the magnetic field is in $z$ direction, then $G_1^0=G_2$, leading to
a further simplification of the matrix elements
Eqs.~(\ref{HMquant1})--(\ref{HMquant5}).

\section{Hund-Mulliken matrix elements, $B\perp x,y$}\label{appendixHMosc}
If the single-electron Hamiltonian is given by Eq.~(\ref{h0})
with a perpendicular field $B\perp x,y$ then we can further
evaluate the integrals Eqs.~(\ref{integral1})--(\ref{integral5}) and the
single-particle matrix elements in
Eqs.~(\ref{matrix})--(\ref{C_offdiag}) as a function of the 
dimensionless inter-dot distance $d=a/a_{\rm B}$ and the magnetic 
compression factors $\alpha_{\pm}(B)=\sqrt{\alpha^2_{0\pm}+B^2/B^2_0}$.  
The single-particle matrix elements are given by
\begin{eqnarray} 
\epsilon_{\pm} &=& \frac{\hbar\omega_z}{2}
\Bigg[1+\frac{3}{16d^2}
+\frac{S}{1-S^2}\left(\frac{\alpha_\pm}{g}+g\alpha_\mp\right.
\nonumber\\
&&\left.\pm\frac{1}{4}\frac{\alpha_{0+}^2-\alpha_{0-}^2}{\alpha_+\alpha_-}\left(g\alpha_\pm-\frac{\alpha_\mp}{g}\right)
\left(1-{\rm erf}(d)\right)\right)\nonumber\\
&& \quad\quad+\frac{S^2}{1-S^2}\left(\frac{3}{4}\left(1+d^2\right)-(\alpha_\pm+\alpha_\mp)\right)\Bigg],\\
t &=& \frac{\hbar\omega_z}{2}\frac{S}{1-S^2}
\bigg[\frac{1}{4}\frac{\alpha^2_{0+}-\alpha^2_{0-}}{\alpha_+\alpha_-}
\left(1-{\rm erf}(d)\right)
\left(\alpha_+ - \alpha_{-}\right)\nonumber\\
&&\hspace{4.7cm}+\frac{3}{4}(1+d^2)\bigg],
\end{eqnarray}
where we have used 
$S=[2\sqrt{\alpha_+\alpha_-}/(\alpha_++\alpha_-)]\exp(-d^2)$.
The (two-particle) Coulomb matrix elements can be expressed as
in Eqs.~(\ref{HMquant1})--(\ref{HMquant5}), where the integrals
Eqs.~(\ref{integral1})--(\ref{integral5}) take the form
\begin{eqnarray}
G_1^\pm &=& \hbar\omega_z 
\frac{2c}{\pi}\frac{\alpha_{\pm}}{\sqrt{1-(2\alpha_{\pm}-1)^2}}\,
{\rm arccos}(2\alpha_{\pm}-1), \\
G_2 &=& G_1^0 = \hbar\omega_z 
\frac{c}{\pi}\frac{(\alpha_++\alpha_-)\,{\rm arccos}(\alpha_++\alpha_--1)}
{\sqrt{1-(\alpha_++\alpha_--1)^2}},\\
G_3 &=& \hbar\omega_z c\sqrt{\mu}
\, e^{2\mu d^2}
\left(1-{\rm erf}(d \sqrt{2\mu})\right),
\label{G3z}\\ 
G_4^\pm &=& \hbar\omega_z c 
\sqrt{\frac{2\alpha_{\pm}(\alpha_++\alpha_-)}{3\alpha_++\alpha_-}}
e^{\mu_{\pm}d^2}
[1-{\rm erf}(d\sqrt{\mu_{\pm}})], 
\end{eqnarray}
where we have introduced 
$\mu=2\alpha_+\alpha_-/(\alpha_++\alpha_-)$ and 
$\mu_{\pm}=\left(\alpha^2_{\pm}+\alpha_+\alpha_-\right)/\left(3\alpha_{\pm}
+\alpha_{\mp}\right)$.
Note that Eq.~(\ref{G3z}) is an approximation which deviates from
the exact result by less than $12$\% in the range $d>0.7$ and $\mu\le 1$
as we have checked by numerical evaluation of the integrals. 

\section{Hund-Mulliken matrix elements, $B\|x$}\label{HMBx}
The Hund-Mulliken calculation for a system of two equal dots with a
magnetic field applied in $x$-direction (Sec.~\ref{parallel}) is 
formally identical to the one with a field in $z$-direction
presented in Sec.~\ref{perpendicular}. For equal dots we set
$\alpha_{0+}=\alpha_{0-}\equiv\alpha_0$,
$\alpha_{+}=\alpha_{-}\equiv\alpha$, and
$\epsilon_+=\epsilon_-\equiv\epsilon$. 
The one-particle matrix elements are then
\begin{eqnarray}
\epsilon &=& \frac{\hbar\omega_z}{2}
\Bigg[\alpha_0+\alpha+\beta+\frac{3}{16d^2\beta^2}
+\frac{S^2}{1-S^2}\frac{3}{4}\left(\frac{1}{\beta}+d^2\right)\nonumber\\
&&\hspace{3cm}
-\frac{S^2}{1-S^2}2d^2\frac{\beta-\alpha}{\alpha}\frac{B^2}{B_0^2}
\Bigg],\\ 
t &=& \frac{\hbar\omega_z}{2}\frac{S}{1-S^2}
\left[\frac{3}{4}\left(\frac{1}{\beta}+d^2\right)
-2d^2\frac{\beta-\alpha}{\alpha}\frac{B^2}{B_0^2}\right]. 
 \end{eqnarray}
Since we consider two equal dots, the matrix elements of the Coulomb
Hamiltonian are formally equal to the matrix elements given in
Ref.~\cite{burkard}, where $F_i$ has to be replaced by $G_i$
defined by
\begin{eqnarray}
G_1&\equiv&G_1^+=G_1^-=G_1^0=\hbar\omega_z\frac{c}{\pi}
\sqrt{\alpha \alpha_0 \beta}\int_0^{\infty}\!\!\!\!dr\, r
\nonumber\\
&&\times\,
{\rm K_0}\left(\frac{\beta r^2}{4}\right)
{\rm I_0}\left(\frac{\alpha-\alpha_0}{4}r^2\right)
e^{-\frac{1}{4}\left(\alpha+\alpha_0-\beta\right)r^2},\\ 
G_2&=&\hbar\omega_z\frac{c}{\pi}\sqrt{\alpha \alpha_0 
\beta}\int_0^{\infty}\!\!\!\!dr\int_{-\infty}^{\infty}\!\!\!\!dz\,\frac{r}{\sqrt{r^2+z^2}}\nonumber\\
&&\quad
\times\,{\rm I_0}\left(\frac{\alpha-\alpha_0}{4}r^2\right)
e^{-\frac{1}{4}\left(\alpha+\alpha_0\right)r^2-\frac{1}{2}\beta (z+2d)^2},\\
G_3&=&\hbar\omega_z\frac{c}{\pi}\sqrt{\alpha \alpha_0 \beta}\,
e^{d^2(B/B_0)^2/\alpha}\nonumber\\
&&\quad\quad\times\int_0^{\infty}\!\!\!\!dr\int_{-\infty}^{\infty}\!\!\!\!dy\,
\frac{r}{\sqrt{r^2+y^2}}\,
{\rm I_0}\left(\frac{\beta-\alpha_0}{4}r^2\right)\nonumber\\
&&\quad\quad\quad\quad
\times e^{-\frac{1}{4}\left(\beta+\alpha_0\right)r^2-\frac{1}{2}\alpha y^2}\cos\left(2yd\frac{B}{B_0}\right), \\
G_4&\equiv&G_4^+=G_4^- =\hbar\omega_z\frac{c}{2\pi^2}\sqrt{\alpha \alpha_0 \beta}\nonumber\\
&&\times\int_{-\infty}^{\infty}\!\!\!\!dy\int_{-\infty}^{\infty}\!\!\!\!dz
\,{\rm K_0}\left(\frac{\alpha_0}{4}\left(y^2+z^2\right)\right)\nonumber\\
&&\times e^{-\frac{1}{4}\left(2\alpha-\alpha_0\right)y^2
-\frac{1}{2}\beta(z-d)^2+\frac{1}{4}\alpha_0z^2}
\cos\left(yd\frac{B}{B_0}\right). 
\end{eqnarray}
Here ${\rm K}_0$ denotes the zeroth order Macdonald 
function and ${\rm I}_0$ is the zeroth order modified Bessel 
function.  The quantities $\alpha$, $\beta$ and $S$ have been
defined earlier.

\section{Heitler-London calculations, $B\parallel x$}\label{appendixBx}
In the following we evaluate the exchange energy $J$  for two coupled
quantum dots in a magnetic field applied perpendicularly to 
the inter-dot axis $({\bf B}\|{\bf x})$ using the Heitler-London approach.
We first study the one-particle problem for an anisotropic quantum dot
with a magnetic field applied perpendicularly to the symmetry axis
of the dot,
\begin{equation}
h^0({\bf r})=\frac{1}{2m}\left({\bf p}-\frac{e}{c}{\bf A}\left({\bf 
r}\right)\right)^2+\frac{m\omega^2_z}{2}\,
\left(\alpha^2_0(x^2+y^2)+z^2\right), 
\end{equation}
where $\alpha_0$ is the ellipticity and ${\bf A}({\bf r})=B(0,-z,y)/2$.
We can separate $h^0({\bf r})=h^0_x(x)+h^0_{yz}(y,z)$ into
a $B$-independent harmonic oscillator
\begin{equation} 
h^0_x(x)=-\frac{\hbar^2}{2m}\frac{\partial^2}{\partial 
x^2}+\frac{m\omega^2_z}{2}\,\alpha^2_0 x^2,
\end{equation}
and a $B$-dependent part
\begin{equation} \label{anisotropic_B}
h^0_{yz}(y,z)=p_y^2+p_z^2-\omega_L L_x
+\frac{m_z\omega^2}{2}\left(\alpha^2\,y^2+\beta^2 z^2\right), 
\end{equation}
with 
$\alpha=\sqrt{\alpha^2_0+(\omega_L/\omega_z)^2}=\sqrt{\alpha^2_0+(B/B_0)^2}$,
and $\beta=\sqrt{1+(\omega_L/\omega_z)^2}=\sqrt{1+(B/B_0)^2}$.
We have not solved Eq.~(\ref{anisotropic_B}) exactly; instead we have
used a variational approach, minimizing the single-particle energy 
\begin{equation}\label{var_energy} 
\epsilon_0=\frac{\langle\psi|h^0_{yz}|\psi\rangle}{\langle\psi|\psi\rangle} 
\end{equation}
as a function of two variational parameters, in order to 
find a good approximate ground state wave function.  A reasonable trial 
wave function $\psi$ should reproduce the anisotropy between $y$ and $z$ in 
the Hamiltonian.  This requirement is fulfilled e.g.  by a Gaussian
\begin{equation} 
\psi_1 (\gamma_1,\gamma_2,y,z)= {\cal N}e^{-\gamma_1y^2-\gamma_2z^2},
\end{equation}
or by mixing Fock-Darwin states $\psi_{0,l}$ with angular momentum
$l=0,2,-2$ and radial quantum number $n=0$, 
\begin{equation}\label{variational}
\psi_2 (\delta_2,\delta_{-2},y,z)= 
\tilde{\cal N}\left[
\psi_{0,0}(y,z)+\sum_{l=\pm 2}\delta_{l}\psi_{0,l}(y,z)\right], 
\end{equation}
where $\delta_{-2},\delta_2$, and $\gamma_1,\gamma_2$ are 
variational parameters and ${\cal N}$, $\tilde{\cal N}$
are normalization constants.
Calculating $\epsilon_0(\gamma_1,\gamma_2)$ and 
$\epsilon_0(\delta_{-2},\delta_2)$, and subsequently minimizing with
respect to the variational parameters, we find that
$\psi_1(m\omega_z\alpha/(2\hbar),m\omega_z\beta/(2\hbar),y,z)$, with the
normalization constant
${\cal N}=(m\omega_z/\pi\hbar)^{1/2}(\alpha\beta)^{1/4}$ is
the best approximate ground-state wave-function in our variational space.
We have also shown 
that including the Fock-Darwin states with angular momentum quantum numbers 
$l=\pm 1$ in $\psi_2$ does not lead to a lower minimum of the energy 
$\langle\psi_2|h^0_{yz}|\psi_2\rangle/\langle\psi_2|\psi_2\rangle$.  The 
full one-particle wave function is then given by
\begin{equation}\label{varfunction} 
\varphi(x,y,z) = \left(\frac{m \omega_z}{\pi\hbar}\right)^{3/4}
\left(\alpha_{0}\alpha\beta \right)^{1/4} 
e^{-\frac{m\omega_z}{2\hbar}
\left(\alpha_{0 } x^2+\alpha y^2+\beta z^2\right)}.  
\end{equation}
Shifting the single-particle orbitals to $(0,0,\pm a)$ in 
the presence of a magnetic field we obtain $\varphi_{\pm a}(x,y,z) = 
\exp(\pm iya/2l_B^2)\varphi(x,y,z\mp a)$.  The phase factor involving the 
magnetic length $l_B=\sqrt{\hbar c/eB}$ is due to the gauge transformation 
${\bf A}_{\pm a}=B(0,-\left(z\mp a\right),y)/2\rightarrow {\bf 
A}=B(0,-z,y)/2$.
Having found an approximative solution for the one-particle problem
in a dot centered at $z=+a$ or $z=-a$, we show that the 
exchange energy is given by Eq.~(\ref{JformalBx})
for a system with two dots of equal size, where $J_0$ denotes the result
from Eq.~(\ref{Jformal}). In the derivation of the formal expression for the
exchange energy $J_0(B,d)$ given in Eq.~(\ref{Jformal}), we have used that
$\varphi_{\pm a}$ was an exact eigenstate of $h^0_{\pm a}$, and therefore
$\langle\varphi_{\mp a}|h^0_{\pm a}|\varphi_{\pm a}\rangle
=S\langle\varphi_{\pm a}|h^0_{\pm a}|\varphi_{\pm a}\rangle$, where
$S=\langle\varphi_{a}|\varphi_{-a}\rangle$ denotes the overlap of the
shifted orbitals.
The approximative solution Eq.~(\ref{varfunction}) for an anisotropic dot
in the presence of an in-plane magnetic field is not an exact eigenstate
of $h^0$. Using the corrected off-diagonal matrix element 
\begin{equation}\label{offdiagonal}
\langle\varphi_{\mp a}|h^0_{\pm a}|\varphi_{\pm a}\rangle
=S\left[\frac{\hbar\omega_z}{2}\left(\alpha_0+\alpha+\beta \right)
+d^2\frac{B^2}{B_0^2}\frac{\beta-\alpha}{\alpha}\right], 
\end{equation}
the result for the exchange energy Eq.~(\ref{JformalBx}) can easily be
derived.


\begin{thebibliography}{99}
\bibitem{footnote1}
For $\hbar\omega_z\approx 20\, {\rm meV}$ and $\alpha_{0\pm}=1/2$ the
approximation is about $5$\% off at zero magnetic 
field, while for $B=B_0\approx 20\,{\rm T}$ the approximation deviates
about $12$\% from the exact value.
\end{thebibliography}
\end{document}